\documentclass[letterpaper,11pt]{article}
\usepackage{natbib}
\usepackage{tabularx} 
\usepackage{amsmath}  
\usepackage{graphicx} 
\usepackage[margin=1in,letterpaper]{geometry} 
\usepackage[final]{hyperref} 
\usepackage{import} 
\usepackage{rotating}
\usepackage{upgreek}
\usepackage{placeins}
\usepackage{siunitx}
\usepackage{booktabs}
\usepackage{physics}
\usepackage{cancel}
\usepackage{amsmath}
\usepackage{acronym}
\usepackage{pdfpages}
\usepackage{bm}
\usepackage{float}
\usepackage{graphicx}
\usepackage{amsmath}
\usepackage{anyfontsize}
\usepackage{gensymb}

\usepackage[font=normalsize]{caption}
\hypersetup{
	colorlinks=true,       
	linkcolor=blue,        
	citecolor=blue,        
	filecolor=magenta,     
	urlcolor=blue         
}
\usepackage{footnote}
\makesavenoteenv{tabular}
\begin{document}

\title{CW Operation of the European XFEL: SC-Gun Injector Optimization, S2E Calculations and SASE Performance}
	\author{Dmitry Bazyl, Ye Chen, Martin Dohlus, Torsten Limberg\footnote{torsten.limberg@desy.de; the authors are listed in alphabetical order.}, \\
	Deutsches Elektronen-Synchrotron DESY, Germany}
\date{29.07.2021}
\maketitle
\begin{abstract}
This note presents optimization results for an injector with a super-conducting gun for the cw operation of the European
XFEL, the corresponding start-to-end simulation calculations for the beam transport to the undulators and SASE calculations of the X-ray intensities for the achievable photon energies.

The optimizations for an injector with a super-conducting gun are done mainly for an already experimentally shown accelerating gradient of 40 MV/m. Different working points with respect to transverse emittance and bunch
length are chosen to see whether the lower transverse emittance of the longer bunches survives the necessary stronger bunch compression to achieve comparable peak currents in the undulators. 
Also, results for
possible further improvement of the gun gradient to 50 MV/m and a modified transverse laser profile (truncated Gaussian) are shown.

The S2E calculations are taking into account all relevant collective effects, including a thorough treatment of the impact of the so called Micro-Bunching
Instability. For that purpose, tracking with the real number of particles is done, with a resolution in space and time fine enough to calculate instability growth model-free.

  
\end{abstract}

\section{Introduction}\label{intro}
Complementary CW operation of the European XFEL is under consideration from the beginning of the project \cite[]{TDR2006}. An additional RF system, an injector with a cw gun and additional cooling power would be the minimum requirements. 
The XFEL in its present pulsed mode has a duty factor of about one percent; keeping all parameters and running in cw mode would increase the necessary cooling by roughly a factor of thirty, since the superconducting linac needs cooling even without RF power.

The normal-conducting l-band gun with its 60 MeV/m gradient already requires extensive water cooling which would be impossible to scale to much higher duty cycle. A cw gun has either to be superconducting or a much bigger copper structure with less gradient, as for example proposed for the LCLS II project.

The present helium cooling plant for the sc linac consumes about 5 MW power in total; scaling that by a factor of thirty is forbidding. In addition, the cooling of the rf modules is limited by the helium piping diameters. A way out is to reduce the accelerating gradient of the rf modules, the thermal load scales quadratic. With a gradient of 7.5 MV/m, corresponding to a final beam energy of about 7 GeV, the necessary cooling flow roughly doubles and can just still be put through the pipes.

The photon energy reach for this electron energy with the present undulators extends to about 6 keV, corresponding to a wave length $\lambda$ of 2{\AA} \cite[]{p_undulator1}. This lowest reachable wavelength is given by undulator and electron beam parameters:
\begin{itemize}
\item{The undulator period $\lambda_u$ and the undulator $K$ value ($K \approx 0.934 \cdot\lambda_u [\text{cm}]\cdot B_0 [\text{T}], B_0$ being the peak magnetic field), which together with the electron beam energy $\gamma$ in units of the electron mass determine the resonant wave length:}

\begin{equation}
\lambda =\lambda_u/{2\gamma^2} (1+ K^2/2)  \label{eqlambda}
\end{equation}

\item{The undulator length $L_u$, which must exceed 20 gain lengths $L_g$ of the exponentially growing SASE pulse energy to provide stable high power photon beams to users. The gain length for a given undulator is governed by electron beam parameters. It increases with transverse emittance $\epsilon_t$ and (slice) energy spread $\sigma_{E}$ and is reduced by higher peak current $I_p$.}
\end{itemize}

So, for a given electron beam energy, undulator parameters $K$ and $\lambda_u$ decide whether a certain wavelength is possible at all while it depends on undulator length and beam parameters $\epsilon_t$, $\sigma_{E}$ and $I_p$ whether the SASE process fits inside the undulator. Gain length calculations in this chapter are carried out using the Ming Xie semi-analytic formula \cite[]{MX1996}.

For the wavelength above, the present undulators ($\lambda_u = 4 \text{cm}$) are at their minimum $K$ (1.6). An emittance of about 1 mm mrad, energy spread of 2.5 MeV and a peak current of 4 kA, rather loose values for the pulsed machine, will be sufficient to reach saturation. 
However, with the rf gradients above, 7.5 MV/m, the final bunch compression stage will be at a beam energy of about 1 GeV. Even at the pulsed XFEL with a beam energy at this stage of 2.5 GeV, the slice emittance starts to increase at final compression due to space charge and coherent synchrotron radiation (CSR) effects \cite[]{Feng2013}. The relative strength of these effects at 1 GeV is increased by a factor 3-6, making the required slice emittances impossible to reach.

Thus, in a more recent paper \cite[]{Brink2014}, the installation of modified RF modules (like for the LCLS II project) with higher cw operation gradients (up to 16 MV/m) up to the last bunch compression chicane is proposed. That provides a beam energy profile throughout the bunch compression section similar to that of the pulsed machine, with comparable beam qualities.
In the following, we consider a scenario with a cw gun, upgraded rf modules throughout the bunch compression system and the present modules in the main linac; the expected final electron beam energy is between 7 and 8 GeV, for this paper we set it to 7.3 GeV (see Fig.1).
 
\begin{figure}[h]

\includegraphics[width=17cm,height=7cm]{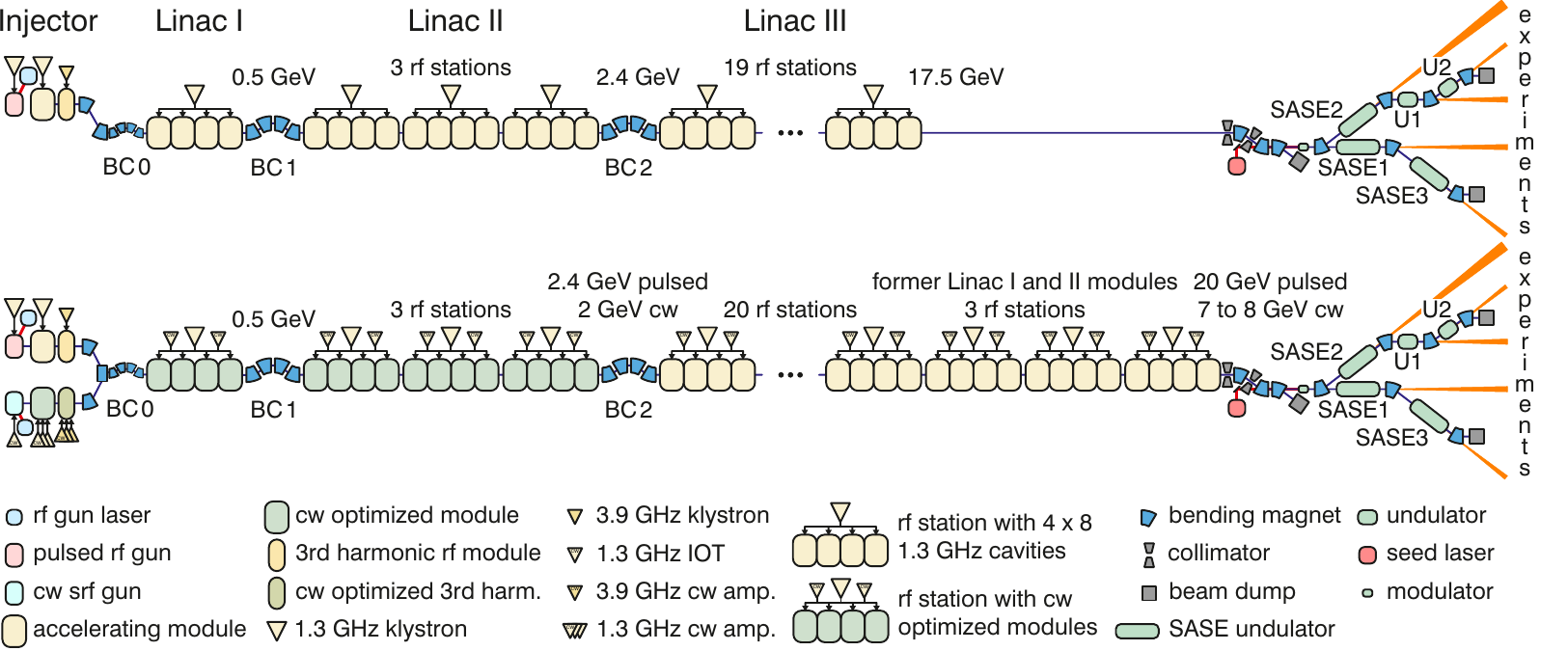}

\caption{European XFEL in cw and pulsed mode}
\end{figure}

To reach photon wavelengths below 1 {\AA} or photon energies around 20 keV, new undulator magnets have to be installed, if lasing on the fundamental mode is aimed for, instead of relying on more advanced schemes like the so called 'third harmonic lasing'  \cite[]{SYHL2012}.

As equ. \ref{eqlambda} suggests, smaller undulator period length results in smaller photon wavelength, e.g., higher photon energy. In Fig, 2,  photon energies over a range of K values are shown for undulators with different $\lambda_u$. For each case, the maximum B-field is varied from 0.2 T to 1.2 T, the operating range of the present undulators. For $\lambda_u = 4$ cm, photon energies for an electron beam energy of 14 GeV, the most common for the pulsed XFEL, is plotted for reference. For the energy of the cw operation, 7.3 GeV, results for three cases of $\lambda_u$ (2,3 and 4 cm) are plotted.


A superconducting undulator (SCU) with  2 cm period has been developed by Karlsruhe Institute of Technology (KIT) and the company Noell GmbH. This device is successfully working at the KIT synchrotron since 2018 \cite[]{Sara}. Similar undulators are manufactured by Noell GmbH and are commercially available. European XFEL is planning an SCU afterburner consisting of 5 modules with a total of 20 m magnetic length and a period between 15 and 18 mm, to extend the photon energy range of SASE2 to higher photon energies \cite[]{Sara1}. 

As mentioned above, the gain length has to obey $L_u \geq 20\cdot L_g$. For a given undulator and given electron energy, it can be kept in check by reducing electron emittance and (slice) energy spread or an increase in peak current. The energy spread, however, is driven by the so called Micro-Bunching-Instability ($\mu$BI). Controlled energy spread increase with a so called laser heater provides stable conditions, but in the end the values downstream of the bunch compression system are considerably higher than the initial spread out of the gun multiplied by the compression factor.

The simulation of beam parameters under consideration of the $\mu$BI is described in chapter \ref{dE}. Since, similar to the SASE process, it starts from noise, the numerical requirements for a proper simulation are very high; particle numbers very close or equal the real particle numbers ($\approx 2\cdot10^9$) have to be tracked.
It turns out that slice energy spread significantly below 2.5 MeV (RMS) as assumed in the calculations for Fig. 2 is hardly possible.

Compressing the bunch to more than 4 kA tends to increase transverse emittance. That leaves reducing the transverse emittance delivered by the injector and preserving it through the transport to the undulator. 

The color coding in Fig. 2 shows necessary emittance values for the different photon energies. For 20 keV at 7.3 GeV electron energy, it has to be around 0.2 $\mu$m. In the following chapters, the path to achieving and transporting these small emittances is laid out.

\begin{figure}[!h]
\centering
\includegraphics[width=10cm]{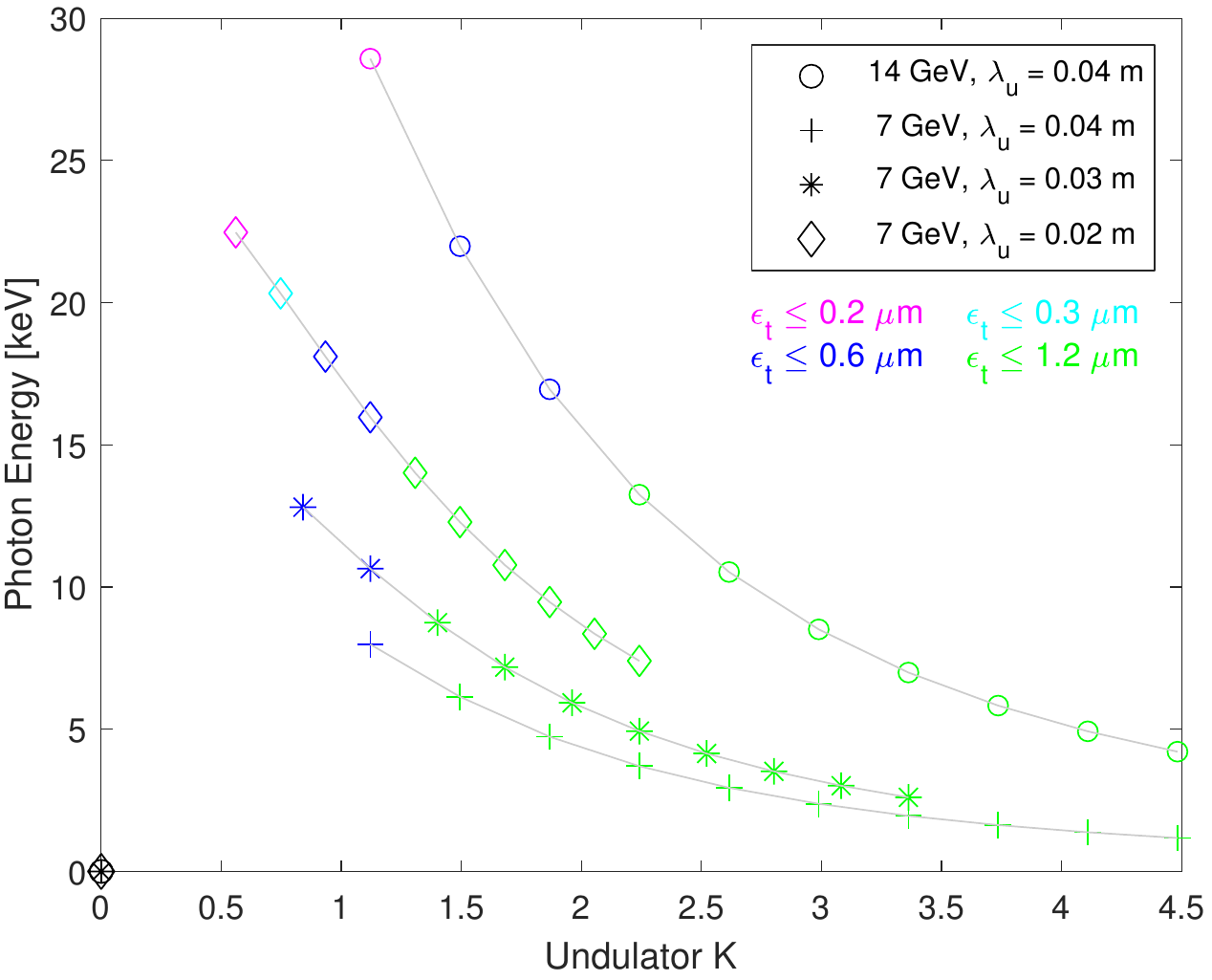}

\caption{Photon energy vs. undulator K for different electron beam energies and different undulator wavelengths. For each undulator the B-field was varied between 0.2 T and 1.2 T. The color code denotes the transverse emittances necessary for a gain length of less than 7 m with a slice energy spread of 2.5 MeV (RMS) and a peak current of 4 kA.}\label{undulatorK1}
\end{figure}

\section{Injector Components and Beam Quality Optimization}
\subsection{Superconducting L-band RF Gun}
An all superconducting L-band RF gun has been proposed in 2005 \cite[]{Seku2005, Seku2007}.  An overview of the recent status of research and development of the L-band SRF gun at DESY can be found in \cite[]{Vogel2018}.

\begin{figure}[htp!]
\centerline{\includegraphics[totalheight=6cm]{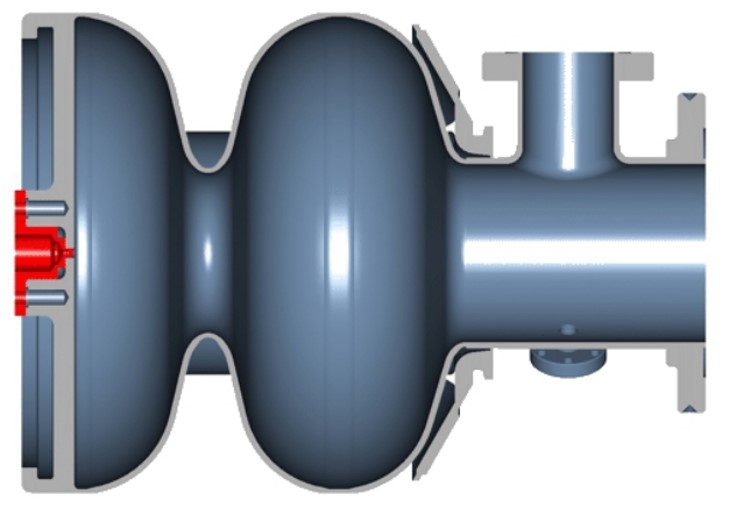}}
    \caption{Layout of the SRF gun with indicated cathode plug (red). } 
    \label{fig:cav3d}
\end{figure}

The geometry of the 1.6-cell SRF gun is based on the TESLA shape (see Fig.  \ref{fig:cav3d} \cite[]{vogel2019}). The
cavity is operated with the accelerating mode $\text{TM}_{010}$ at the resonance frequency of 1.3 GHz. The cavity is made of bulk
niobium (Nb) while lead (Pb) is implemented as the cathode material. The
cathode is inserted via the cathode plug and its position can be
adjusted such that the focusing RF field in the vicinity of the cathode is
optimized.  Details concerning
properties of lead cathodes related to photo-emission are given in Section
2.2.

\begin{figure}[htp!]
\centerline{\includegraphics[totalheight=7cm]{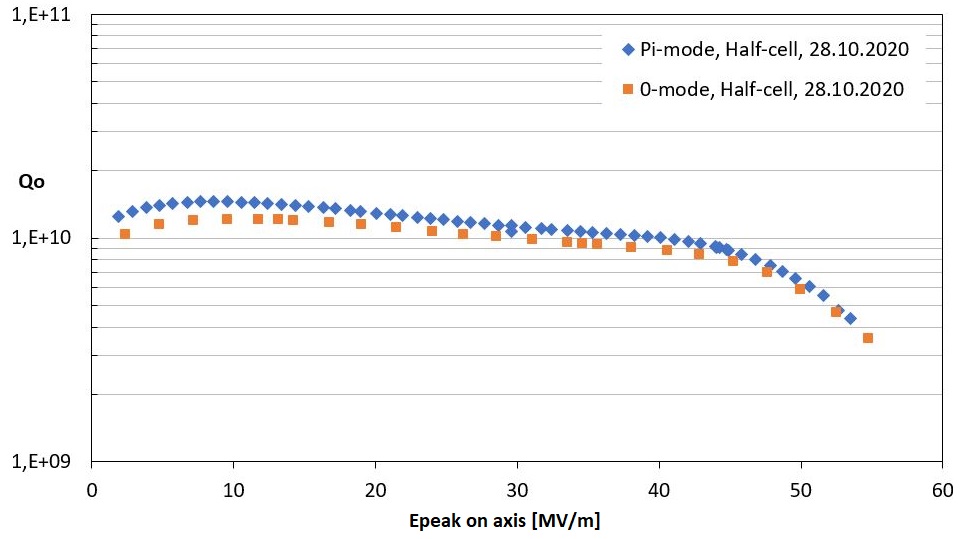}}
    \caption{Results of recent vertical tests of DESY SRF L-band gun carried out at KEK. Courtesy: The DESY
SRF photo injector team and KEK(Japan).} 
    \label{fig:kek}
\end{figure}

One of the key characteristics of the SRF gun is the peak electric field on axis.
The fundamental limitation of the maximal peak electric field on axis for
cavities made of bulk Nb is characterized by the critical magnetic field at the
inner surface of 200 mT. Below the fundamental limit the maximal
peak electric field is dictated by the quality of the inner surface of the
SRF gun. The DESY L-band SRF gun has a design value for the peak electric field on axis of 40 MV/m.
Vertical cold tests of early gun prototypes indicated
the possibility of achieving the peak electric field on axis of up two 60 MV/m. 
Most recent vertical cold test which has been prepared and carried out by the SRF group at KEK (the collaborator partner of DESY concerning SRF gun development) yield
the peak electric field on axis of up to 50 MV/m without field emission (see Fig. \ref{fig:kek})\cite[]{KEK}.
As it is shown further down,  higher peak electric field on axis yields lower transverse slice emittance.

\subsection{Properties of lead cathode}

The electron emission process from metallic photo cathodes is elucidated in \cite[]{SLAC}. The root mean square (rms) normalized emittance can be typically expressed as

\begin{equation}
    \epsilon_{n} = \sigma_x\sqrt{\frac{\hbar\omega - \phi_{eff}}{3mc^2}},\label{eq:thermal}
\end{equation}
where $\sigma_x$ stands for the rms size of the laser spot on the cathode surface and the rest of Eq.\ (\ref{eq:thermal}) represents the dimensionless transverse momentum of photo-emission. The terms $\hbar\omega$, $\phi_{eff}$, $m$ and $c$ are denoted as the photon energy, material effective work function, electron mass and the light speed in vacuum, respectively. For convenience, Eq.\ (\ref{eq:thermal}) is further expressed as $\epsilon_{n} =\sigma_x\sqrt{2E_{th.kin.}/{3mc^2}}$, where the term $E_{th. kin.}$ is used to define the thermal energy of the cathode in the ASTRA simulations.

Although an exact prediction of material work function is quite challenging, it is still possible to summarize the state of the art based on literature studies, for instance, more insights about lead photo cathodes are reported in \cite[]{BNL} and \cite[]{HZB}. A summary of available data is given in Table \ref{tab:thermal}. As an example, the numbers in last columns indicate the intrinsic emittance for an rms laser spot size of 0.25 mm (1 mm in diameter for a uniform distribution). To generate 100 pC with a Quantum Efficiency (QE) of 0.9E-4 would require a laser (single) pulse energy of about 5.35 $\mu$J. This requirement is achievable with modern laser systems. From this point of view, the demonstration at HZB (3rd row) would suggest an estimation of the thermal emittance to be 0.121 $\mu$m for a 1 mm spot (if no surface potential reduction considered). As one adjusts the spot size, a calculation should be done according to Eq.\ (\ref{eq:thermal}). Note in addition, that other photo cathode associated effects and their potential contributions to QE and / or thermal emittance, such as laser pulse heating, surface roughness, etc., will be studied in a later stage.

\begin{table}[!htbp]
\begin{minipage}{\textwidth}
	\centering
	\setlength{\tabcolsep}{20pt}
	\caption{Available data of lead photo cathodes and estimations.}
	\begin{tabular}{lcccccc}
		\hline\hline
		&$\hbar\omega$ &$\Phi$&$\hbar\omega$-$\Phi$ &$\textnormal{QE}$ &$\sigma_{x}$&$\epsilon_{n}$\footnote{for a laser spot size of 1 mm in diameter, with homogeneous distribution.}\\
		&eV&eV&eV&-&mm&$\mu$m\\
		\hline
		{\bf 193 nm\footnote{BNL, tests on samples, no typical cavity treatments, 1 MV/m.}}	&6.42   &3.88  &2.54   &5.41E-3    &0.25 &0.322\\
		{\bf 213 nm\footnote{BNL, tests on samples, no typical cavity treatments, 1 MV/m.}}	&5.82   &3.88 &1.94    &2.72E-3    &0.25 &0.281\\
		{\bf 258 nm\footnote{HZB, realistic SRF cavity environment, no Schottky.}}	&4.81   &4.45  &0.36   &0.90E-4    &0.25 &0.121\\
		{\bf 258 nm\footnote{HZB, realistic SRF cavity environment, a Schottky reduction of 0.08 eV for about 4.8 MV/m.}}	&4.81   &4.37  &0.44   &0.95E-4    &0.25 &0.134\\
		\hline\hline
	\end{tabular}\label{tab:thermal}
\end{minipage}
\end{table}

\subsection{Layout of the CW injector and parameters for optimization}
The layout of the CW injector with the length of 15 m for the beam quality optimization consists of the DESY SRF L-band gun,  superconducting focusing solenoid and one accelerating module which is formed by eight 1.3 GHz SRF
TESLA cavities with the maximal peak electric field on axis of 32 MV/m. The electric field
profile of the fundamental mode of the SRF gun is evaluated using CST
MWS\textregistered~\cite[]{cst}.  Parameters which are used for the optimization of  beam dynamics in the CW injector
are summarised in Table \ref{Tab:optpar}.  The RF phase of the accelerating module is set on
crest and optimized further on during the compression stage in start-to-end simulations. The
amplitude of the electric field of the last four cavities is set to the maximum value of 32 MV/m which corresponds to the accelerating gradient of 16 MV/m.

\begin{table}[htp!]
\centering
\begin{tabular}{@{}lc@{}}
\toprule
\textbf{Definition}                & \multicolumn{1}{l}{\textbf{Variable}} \\ \midrule
rms laser spot size [mm]               &   $\sigma_x, \sigma_y$                     \\
rms laser pulse lenght [ps]                     & $\sigma_z$                                 \\
gun phase offset [\degree]        & $\phi$                                  \\
solenoid position [m]                   & $S_\text{pos}$                                  \\
peak magnetic field on axis of the solenoid [mT] & $B_{\text{p}}$                               \\
position of the first accelerating module [m]& $C_{\text{pos}}$                               \\
peak electric field of the first four cavities in the first module [MV/m]& $E_{\text{1}}$                               \\\bottomrule
\end{tabular}
\caption{Variables used for multi-objective optimization of the CW injector.}
\label{Tab:optpar}
\end{table}

\subsection{Multi-objective optimization of the CW injector}
The optimization of beam dynamics in the injector has been performed for two
objective functions: the rms bunch length and the rms 
transverse projected emittance.  For the optimization purposes we refer to C++
code written in LBNL \cite[]{Popad14} while ASTRA \cite[]{astra} is implemented for beam dynamics simulations.  The number of particles in ASTRA simulations used for the optimization is 10 000.  In order to improve the accuracy of simulations,  interesting scenarios are further recalculated with higher mesh density and 200 000  particles. The C++
code is based on the NSGA-II algorithm \cite[]{nsga2}. A single optimization takes approximately 24 hours using 80 CPUs on the DESY cluster. The result of the
optimization is a Pareto frontier which contains non-dominated solutions.

Electron bunches with the charge of 100 pC are shaped
by two combinations of the laser profiles.  First, is the
combination of the longitudinal Gaussian and the transverse radial uniform
laser profile. Second is the longitudinal flat top and the transverse
truncated-Gaussian at 1$\sigma$.  

\subsection{Optimization Results}
The aim of this section is to discuss results of the multi-objective optimization which has been carried out for the CW SRF photo-injector base on the L-band SRF gun. The goal of the optimization is to minimize transverse emittance and the longitudinal length of bunches at the injector exit. First, positioning of the solenoid is discussed. Further on, for given position of the solenoid optimization is carried out for two combinations of laser shapes (longitudinal Gaussian and the transverse radial uniform
laser profile; longitudinal flat top and the transverse
truncated-Gaussian at 1$\sigma$). Based on the obtained data, a number of bunches has been selected and studied further in start-to-end simulations.

\subsubsection{Position of the Focusing Solenoid}
\begin{figure}[htp!]
\centerline{\includegraphics[totalheight=6cm]{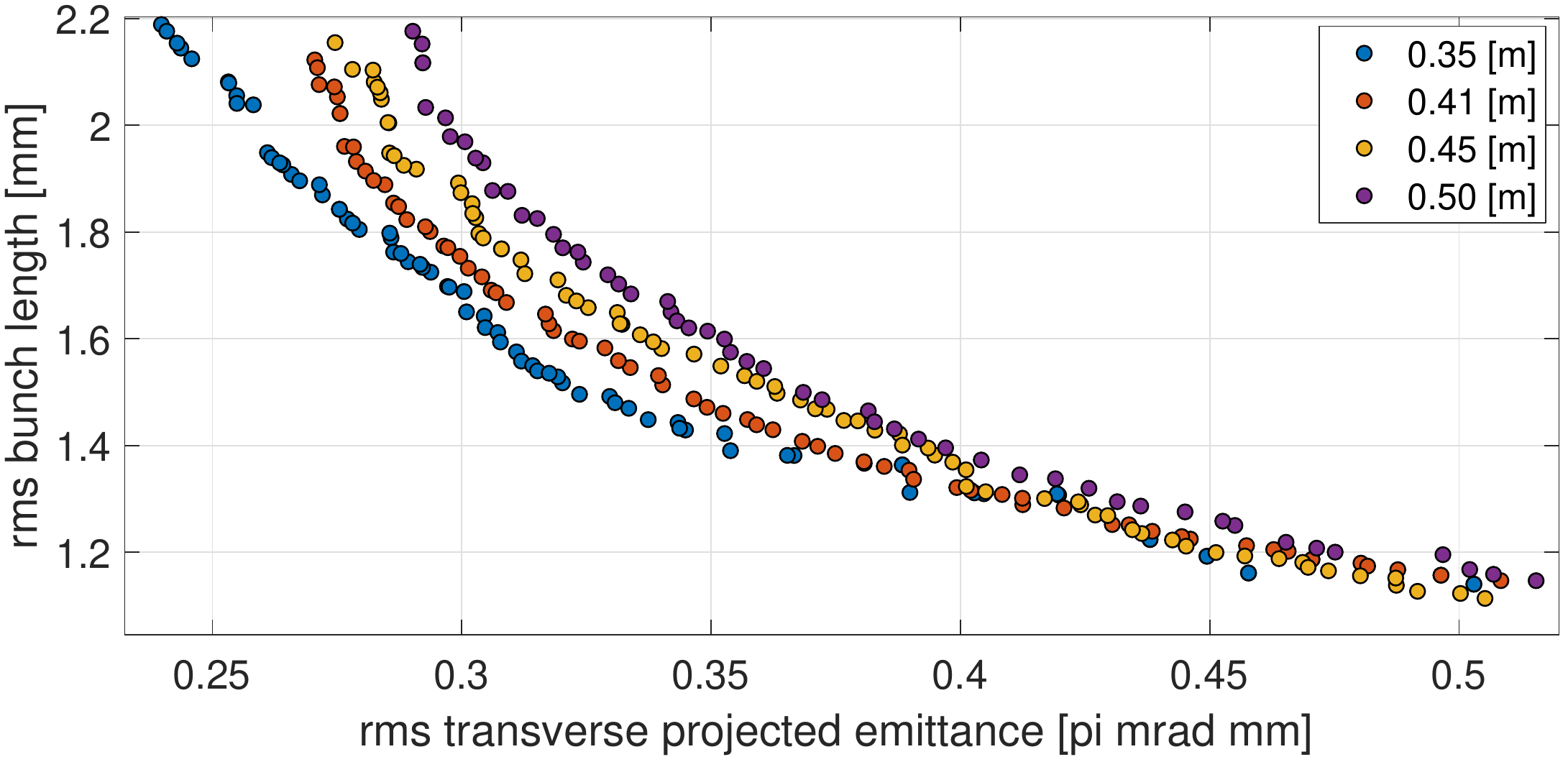}}
    \caption{Results of the optimization obtained for different positions of the solenoid.  Peak electric field in the SRF gun $E_\text{p}$ = 40 $MV/m$; $E_{\text{th. kin.}} $= 0.22 eV; cathode is retracted from the back wall by 450 $\mu$m. Laser profile: longitudinal Gaussian, transverse radial uniform.}  
    \label{fig:solpos}
\end{figure}

The position of the solenoid is defined as the distance between the cathode and the center of the solenoid.  In general, independent of the value of the peak electric field in the gun, allocating the focusing solenoid closer to the cathode plane yields better results in terms of the transverse projected
emittance.  In addition to spatial constrains,  allocating the solenoid too close to the SRF gun might
perturb the superconducting state of the cavity.  The final position of the focusing solenoid will be defined based on geometric constraints (including magnetic shielding) in the cooling cryostat which will host the SRF gun and the superconducting solenoid. The cryostat is currently at the design stage.  Presently, the foreseen position of the solenoid is within a range from 0.41 m up to 0.5 m. Figure \ref{fig:solpos} indicates Pareto fronts obtained for different positions of the solenoid with a peak electric field in the gun of 40 MV/m and bunches generated by longitudinal Gaussian and transverse radial uniform laser profile.  Thermal emittance is characterized by a moderate value of $E_{\text{th. kin.}} $ = 0.22 eV (corresponds to 0.5 $\mu$m/mm of thermal emittance). For further optimization runs we set the solenoid position at 0.41 m.

\subsubsection{Impact of the Laser Profile and the Peak Electric Field in the Gun on the Transverse Emittance}
\begin{figure}[htp!]
\centerline{\includegraphics[totalheight=5.35cm]{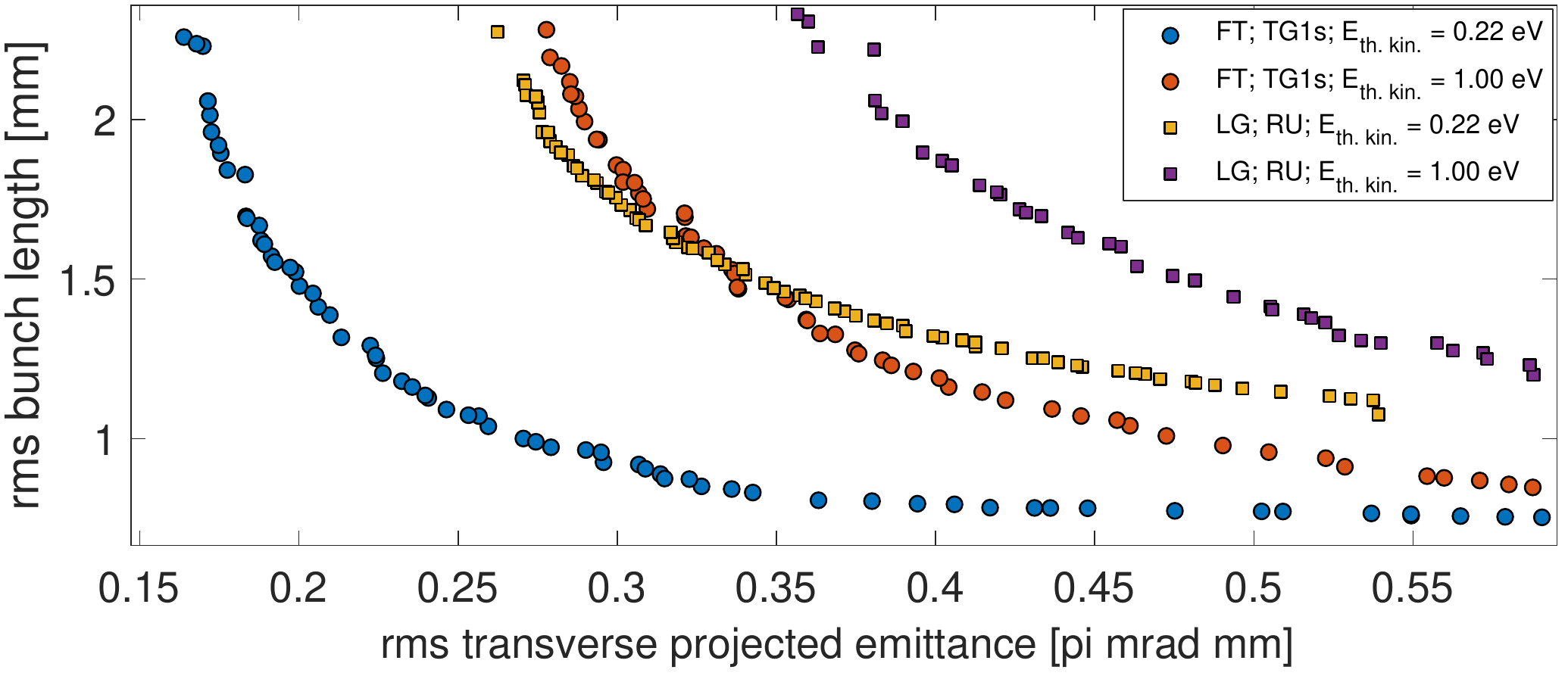}}
    \caption{Results of the optimization for the peak electric field on axis of the SRF gun of 40 MV/m.  Solenoid is located at 0.41 m; cathode is retracted from the back wall by 450 $\mu$m. Laser profiles:  longitudinal flat top (FT), transverse truncated-Gaussian (at 1$\sigma$) (TG1s);longitudinal Gaussian (LG), transverse radial uniform (RU).}  
    \label{fig:40new}
\end{figure}

\begin{figure}[htp!]
\centerline{\includegraphics[totalheight=5.35cm]{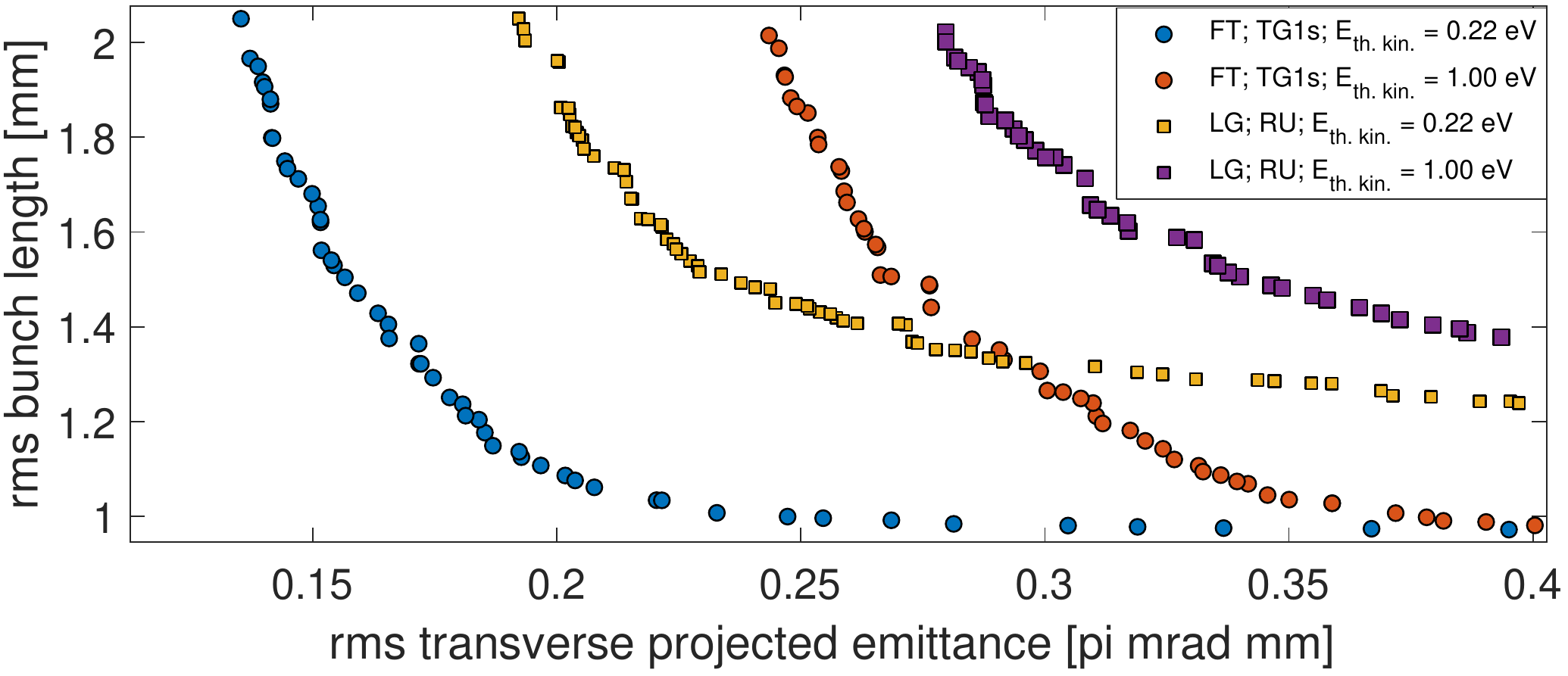}}
    \caption{Results of the optimization for the peak electric field on axis of the SRF gun of 50 MV/m.  Solenoid is located at 0.41 m; cathode is retracted from the back wall by 450 $\mu$m. Laser profiles:  longitudinal flat top (FT), transverse truncated-Gaussian (at 1$\sigma$) (TG1s);longitudinal Gaussian (LG), transverse radial uniform (RU)} 
    \label{fig:50mv}
\end{figure}

\begin{figure}[htp!]
\centerline{\includegraphics[totalheight=5.4cm]{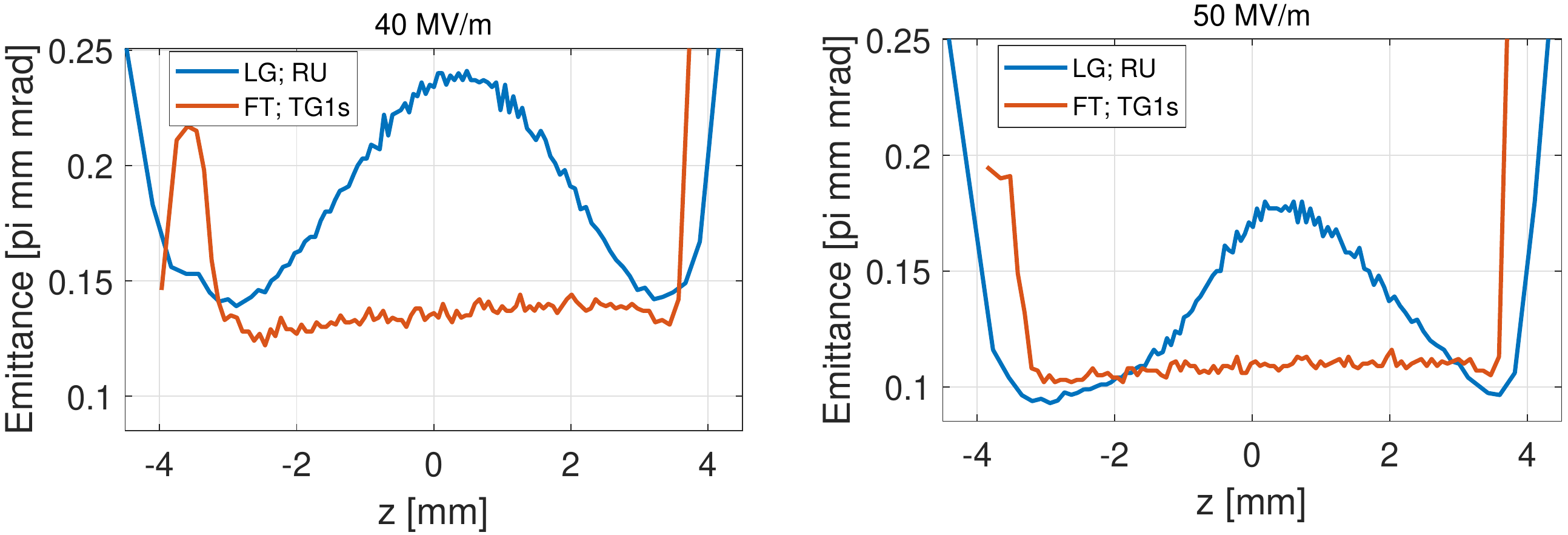}}
    \caption{Comparison of transverse slice emittance of bunch distributions formed by two combinations of laser profiles. Laser profiles: longitudinal flat top (FT), transverse truncated-Gaussian (at 1$\sigma$) (TG1s);longitudinal Gaussian (LG), transverse radial uniform (RU). Solenoid position 0.41 m. $E_{\text{th. kin.}} $= 0.22 eV.} 
    \label{fig:slice_40_50}
\end{figure}

In the previous sub-section we considered laser shaping with longitudinal Gaussian and transverse radial uniform distribution. In order to further minimize the transverse emittance we refer to more advanced laser shaping and further on compare the results. As reported in \cite[]{tgs1}, transverse truncated-Gaussian laser shaping can reduce transverse slice emittance by up 25 \% in comparison to radial uniform. Presently, European XFEL in pulsed mode is operated with the transverse laser profile which is very similar to the truncated-Gaussian at 1$\sigma$ \cite[]{lutzlaser}.  In addition, longitudinal flat top laser profile allows to minimize the transverse projected emittance. Figure \ref{fig:40new} illustrates results of the injector optimization when the solenoid position is defined at 0.41 m  with reasonable value of the peak electric field of 40 MV/m in the SRF gun. Results correspond to two values of thermal emittance which are covering the foreseen range for the lead cathode.  Figure \ref{fig:50mv} illustrates  optimization results similar to what has been shown in Fig. \ref{fig:40new} but this time the peak electric field in the SRF gun is set to more aggressive value of 50 MV/m. Comparison of the transverse slice emittance for the two considered laser shapes is shown in Fig. \ref{fig:slice_40_50} (left - peak electric field in the gun is set to 40 MV/m; right - 50 MV/m). In this comparison the thermal emittance corresponds to a moderate value  $E_\text{th. kin}$ of 0.22 eV.  The bunch length is similar for each peak electric field in the gun. 

The implementation of the transverse truncated Gaussian laser profile brings significant benefits in terms of minimization of the transverse slice emittance. It is also clear that higher peak electric field on axis of the RF gun (i.e. higher peak electric field at the cathode during emission process) allows to further minimize transverse slice emittance. 

\FloatBarrier

\subsubsection{Selection and Description of Bunch Distributions for Start-to-End simulations}
\begin{figure}[htp!]
\centerline{\includegraphics[totalheight=5.3cm]{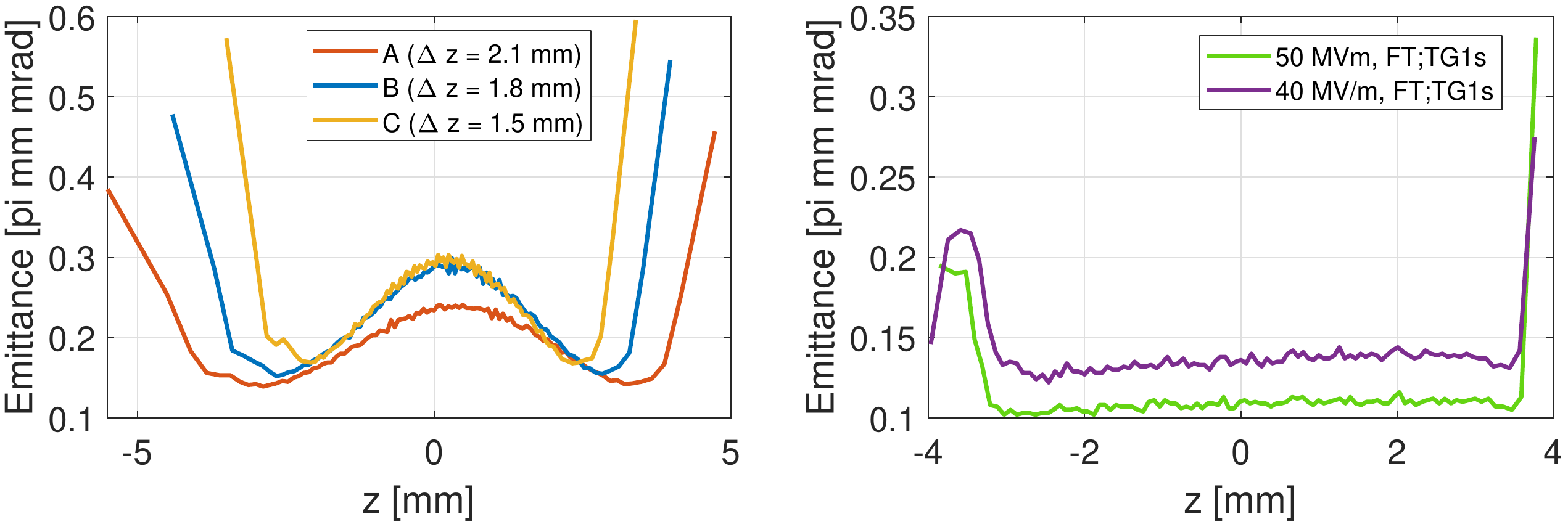}}
    \caption{Transverse slice emittance of bunch distributions selected for further start-to-end simulations. Left: bunches with varying bunch length. Laser profile LG;RU. Solenoid position 0.41 m. Initial thermal kinetic energy $E_{\text{th. kin.}} $= 0.22 eV. Right: bunches with lowest transverse slice emittance. Solenoid position 0.41 m. Initial thermal kinetic energy $E_{\text{th. kin.}} $= 0.22 eV } 
    \label{fig:slice_s2e}
\end{figure}

Based on the results of the injector optimization described in the previous section, five bunch distributions were selected to further study their properties in start-to-end simulations. The first three distributions are generated by the transverse radial uniform and longitudinal Gaussian laser and differ from each other by varying the bunch length $\Delta$z (see Fig. \ref{fig:slice_s2e} (left)). The motivation is to study the trade-off between higher peak current and lower transverse slice emittance. The remaining two bunch distributions are characterized by the lowest values of the transverse slice emittance among bunch distributions obtained during optimization of the CW XFEL injector for each considered peak electric field on axis in the gun (see Fig. \ref{fig:slice_s2e} (right)).  Table \ref{Tab:abctab} summarizes resulting bunch properties as well as details related to the considered setup of the laser and the gun in each particular case.

\begin{table}[htp!]
\centering
\begin{tabular}{lccccc}
\toprule
\multicolumn{1}{c}{\textbf{Parameter}}          & \textbf{A} & \textbf{B} & \textbf{C} &                                & \multicolumn{1}{l}{}                               \\ \hline
longitudinal laser profile                      & \multicolumn{3}{c}{Gaussian}         & \multicolumn{2}{c}{flat top}                                                                   \\
transverse laser profile                        & \multicolumn{3}{c}{radial uniform}   & \multicolumn{2}{c}{\begin{tabular}[c]{@{}c@{}}truncated-Gaussian\\  (at 1 sigma)\end{tabular}} \\
peak electric field in the gun                  & \multicolumn{3}{c}{40 MV/m}          & 40 MV/m                                  & 50 MV/m                                             \\
thermal emittance {[}pi mm mrad{]}              & 0.14      & 0.16       & 0.17       &    0.13                                      & 0.10                                            \\
rms laser spot size {[}mm{]}                    & 0.26       & 0.29       & 0.31       &    0.23                             & 0.19                                              \\
rms laser pulse length {[}mm{]}                 & 8.70       & 7.30       & 5.60        & 8.29                                         & 8.30                                                 \\
transverse. proj. emitt. (x/y) {[}pi mm mrad{]} & 0.32       & 0.35       & 0.38       &     0.15                                  & 0.12                                                \\
rms bunch length {[}mm{]}                       & 2.13       & 1.80       & 1.53       &     2.04                                     & 2.05                                                                      \\\bottomrule
\end{tabular}
\caption{Parameters of bunches seleceted for start-to-end simulations and parameters related to the considered setup of the laser and the gun. Transverse projected emittance and rms bunch length values correspond to bunch position at 15 m distance from the cathode}
\label{Tab:abctab}
\end{table}
\FloatBarrier

\section{Start-to-End Simulations}
The optimized electron bunches from the CW injector are evaluated through the start-to-end (S2E) beam dynamics simulations of the conceptual CW XFEL. This is done by tracking selected bunches from the CW injector further downstream to the entrance of the undulator beam line with all collective effects on. The bunch qualities, characterized by the peak current, slice energy spread, slice emittance, etc. are then assessed in case studies. First SASE simulations are also performed at this stage to explore the lasing performance in hard X-ray regime to validate analytical calculations performed in Sec.\ \ref{intro}.

A more specific goal of S2E beam dynamics optimization is set such that an optimized 100 pC bunch can be compressed to a peak current of kAs with sufficient charge occupation around the current peak while the corresponding slice emittance and energy spread can be kept reasonably small. Since the overall compression and thus the final bunch length is sensitive to rf parameters (i.e.\ amplitudes and phases of acceleration modules A1, AH1, L1 and L2), the optimization approach needs to consider both multiple longitudinal beam dynamics parameters and the rf parameters. More importantly, the mapping between the two should be properly taken into account. A general optimization procedure including the mapping between longitudinal beam dynamics parameters and rf parameters as well as an iterative algorithm used to search for the rf parameters are presented in \cite[]{Zagorodnov-PRAB2019}. Another key factor to the overall compression scheme is appropriate determination of the energy spread of the compressed bunches in light of the micro-bunching effect. In this work this is realized by high-resolution numerical simulations in the code IMPACT-Z \cite[]{impactz1} and \cite[]{impactz2}. A large number of macro-particles, e.g. a few tens (/hundreds) of millions, as well as very fine longitudinal mesh resolution, are adopted for simulating the 100 pC bunches. Note, in addition, that for simplicity this conceptual design follows a general layout of the current XFEL machine for the time being. At this stage, no particular designs of the optics or significant modifications to the beam line are considered. Interested readers are also referred to \cite[]{XFEL} for a detailed description of the current XFEL facility and to \cite[]{Compression2} and \cite[]{Compression3} for previous bunch compression studies at the XFEL.

This section is organized as follows. Section \ref{dE} presents dedicated high-resolution micro-bunching studies for compressing the bunches to kAs' peak currents after the last bunch compression stage. In Section \ref{tracking}, the S2E simulation results are shown before the undulator beam line for the selected bunches from the injector optimization, as shown in Table \ref{Tab:abctab}. In Section \ref{SASE}, SASE simulations are performed to explore the lasing performance in sub-nanometer and sub-angstrom regimes based on the bunches obtained in Sec. \ref{tracking}, further confirming theoretical predictions as presented in Sec.\ \ref{intro}.

\subsection{Micro-Bunching Instability Studies}\label{dE}

\subsubsection{Introduction to the Micro-Bunching Instability}\label{IuB}

Multi-stage bunch compression systems, like the one in the European XFEL, are successions of straight sections with accelerating rf elements and magnet chicanes. 
In the straights, current modulations of the electron bunch cause energy modulations via the impedances of space charge and wake fields, which in the next magnet chicane are then converted into stronger current 
modulations and so forth. Of course, the CSR impedance also contributes to the amplification.
The gain of that process at modulation wave lengths small compared to the bunch length can be orders of magnitude. The instability is damped by slice energy spread, which can be enhanced by a so-called laser heater \cite[]{LH1}.

At the electron gun exit, the initial current modulation stems from shot noise and, additionally, from non-flatness of the laser pulse illuminating the cathode. Thus, modelling of the micro-bunching instability requires tracking very high particle numbers close to the actual amount of particles ($6\cdot 10^8$ for 0.1 nC bunch charge) from the very beginning with sufficient longitudinal resolution to resolve the impedance spectrum.

For the first about three meters from the cathode to the drift after the gun cavity the SC codes Astra (r-z mesh, \cite[]{astra} and FMM (mesh-less fast multipole method, \cite[]{FMM}) are used. The remaining injector and bunch compression system to the exit of the last bunch compressor (BC2) is simulated with the code Impact-Z. For the preparation of the particle distribution after the gun calculation we used two methods:

\begin{enumerate}
\item{
A simulation with few macro particles (about 100 electrons per macro particle) followed by a method to increase the particle number, and}
\item{the combination of many particle distributions obtained by many independent gun simulations.}
\end{enumerate}

For both approaches macro particle sets with one or few electrons per macro particle are generated as Impact-Z input. The validation of (1) is obtained by tests with increased number of macro particles and by comparison with (2). The reasoning for (2) is linearity due to small noise amplitudes (as micro bunching effects on the first meters are very weak).

\begin{figure}[h!]
\centering
\includegraphics[width=14cm,height=9cm]{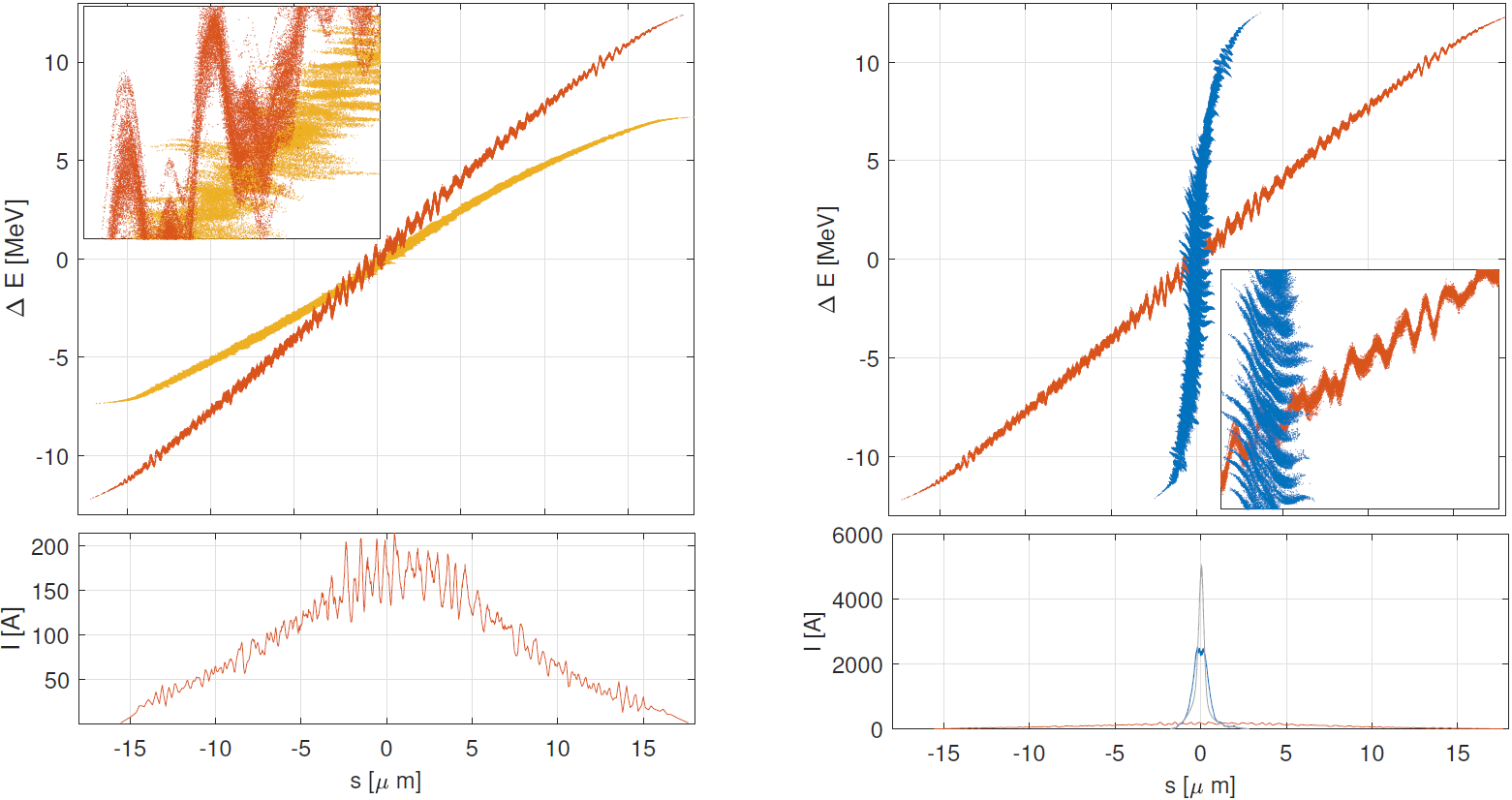}
\caption{Longitudinal phase space and current profile. Left:  After BC1 (yellow) and before BC2 (red), right:  Before (red) and after BC2 (blue). Grey current profile: reference with laser heater on.}\label{MDuB1}
\end{figure}

\begin{figure}[h!]
\centering
\includegraphics[width=14cm,height=9cm]{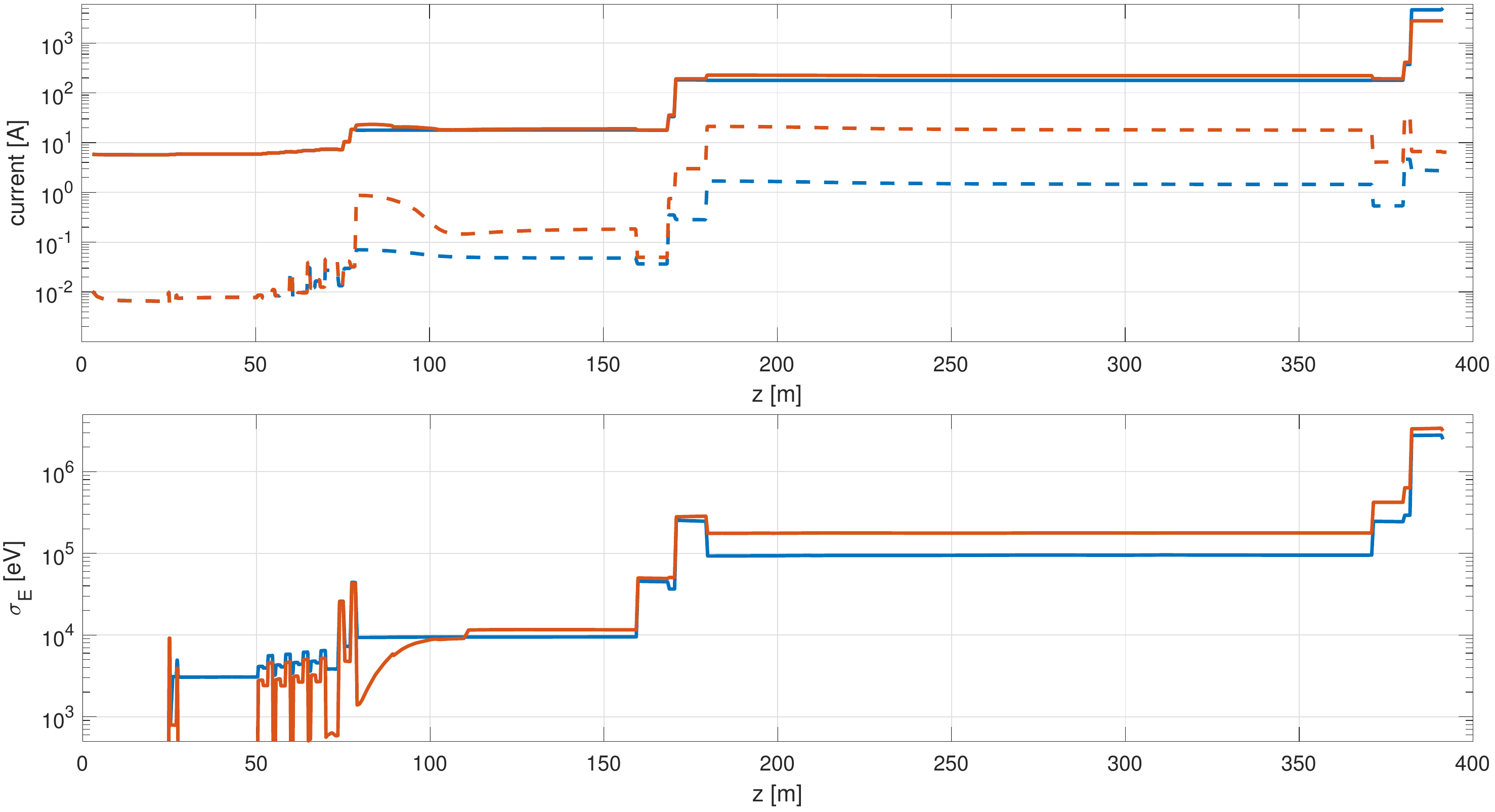}
\caption{Simulated peak current, RMS intensity jitter around the peak current point, and slice energy spread along the bunch compression system with (blue lines) and without (red lines) laser heater}\label{MDsE}
\end{figure}

For demonstration we show in Figs. \ref{MDuB1}-\ref{s2e_uB2} simulation calculation results with the code Impact-Z for 40 MV/m gradient and radial uniform laser at the gun (case B). Similar simulations have been done for the other cases.

The gun-laser profile is perfect without any ripple so that the instability starts at the cathode from white shot noise.
 The evolution of the longitudinal phase space in the last compression stage is shown. The magnet chicanes of the three-stage bunch compression system in the European XFEL are, for historical reasons, called BC0, BC1 and BC2. 

The left-hand plot shows that the current modulation of the bunch causes an energy modulation on its travel from BC1 to BC2. In the right-hand plot, the final longitudinal compression is shown. The phase space distribution after BC2 is blown up in the longitudinal direction at locations where the gradient of the incoming energy modulation is either over- or de-compressing the particle distribution. As a result, the achievable peak current is limited to about 2 kA.

A close look at the longitudinal phase space after BC1 shows a very high frequency longitudinal blow up of lesser magnitude caused by the instability of the previous compression stage, illustrating the need for huge particle numbers for these simulations.

The laser heater at the E-XFEL is an undulator of about one-meter length in the middle of a magnet chicane, where the electron bunch travels together with a high-power laser beam. The energy changes of the particles induced by the electric field of the laser together with the movement of the electrons through the chicane effectively increases the incoherent slice energy spread []. 

In Fig. \ref{MDsE}, the initial slice energy spread is 0.5 keV. The longitudinal bunch compression factor is about 800, so without instability or laser heater the slice energy spread after compression would be about 0.4 MeV.
 
Instead, it increases, driven by the micro-bunching instability, to a final value of 3.5 MeV (red line).  If the laser heater is on, it increases the slice energy spread to about 3 keV (blue curve @ z=30m), the final value is around 2.5 MeV, close to what applying the  bunch compression factor would yield. 

Furthermore, by preventing the energy and intensity modulations and the resulting longitudinal phase space blow-ups described above, the damping provided by the laser heater makes a peak current of 5 kA possible (blue curve @z=380m in Fig. \ref{MDsE}, grey current profile in Fig. \ref{MDuB1}).

\subsubsection{Simulation Results}

\begin{figure}[!h]
\centering
\includegraphics[width=12cm,height=8cm]{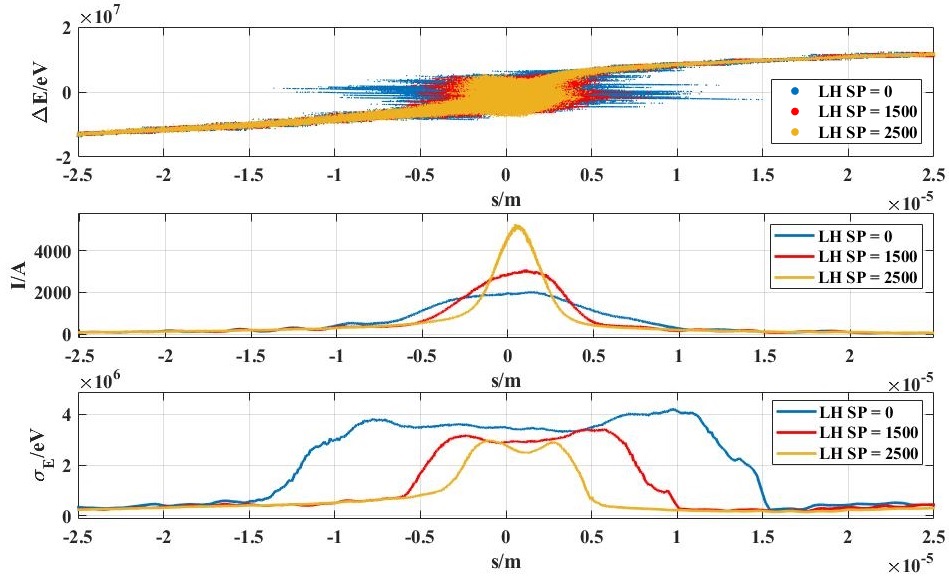}
\caption{Simulated longitudinal phase spaces after the last bunch compression stage and corresponding bunch quality parameters for different set-points of the laser heater model.}\label{s2e_uB1}
\end{figure}

\begin{figure}[!h]
\centering
 \includegraphics[width=8cm,height=9cm]{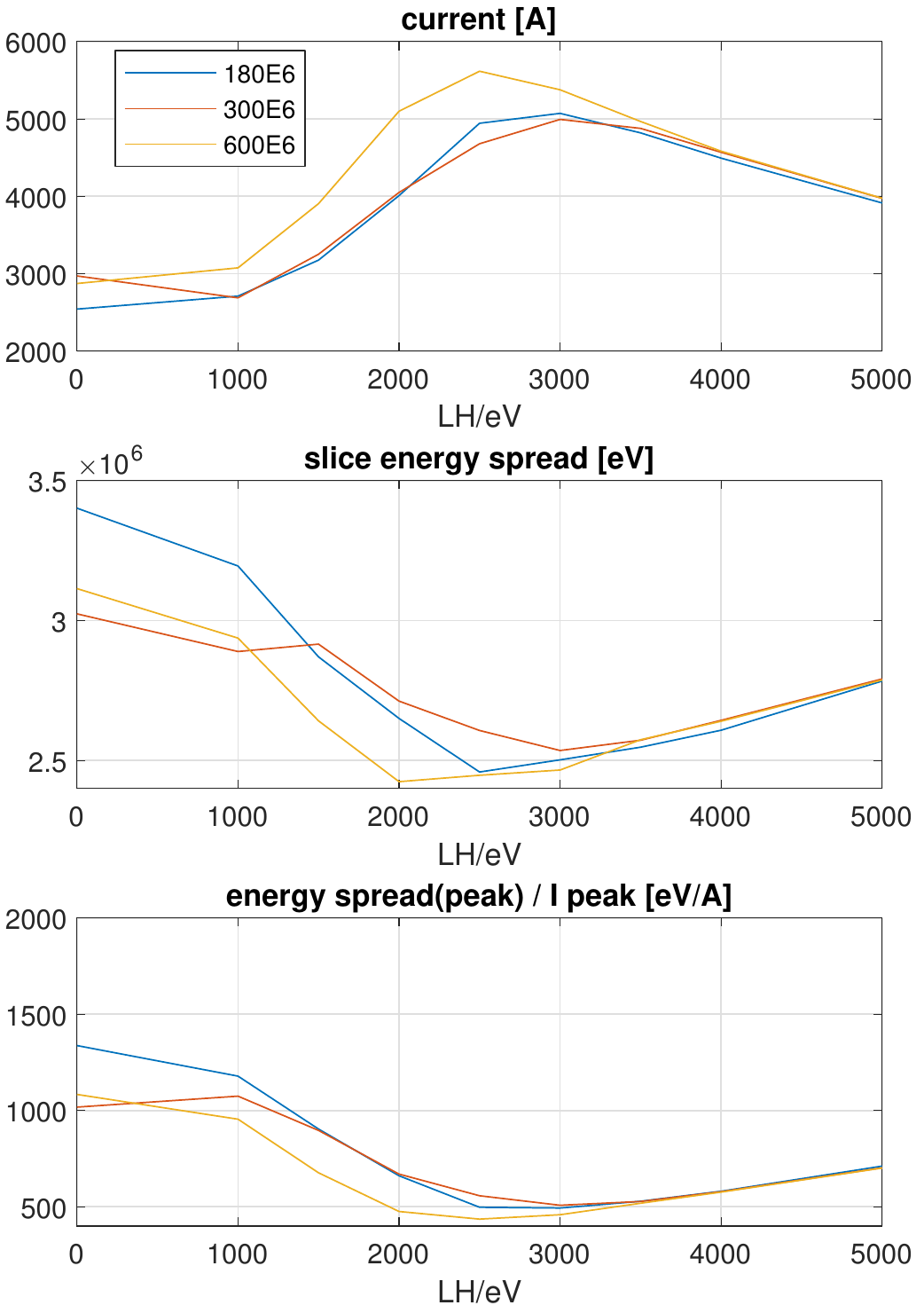}
\caption{Scanning the laser heater set-point vs. achievable peak current, central slice energy spread and their ratio after the last bunch compression stage for different particle numbers tracked.}\label{s2e_uB2}
\end{figure}

The simulation results are summarized in Figs.\ \ref{s2e_uB1} and \ref{s2e_uB2}. A series of simulated longitudinal phase spaces after the last bunch compressor, as well as the key beam properties along the bunches for various set-points of the laser heater model are shown. As illustrated, increasing the laser heater set-point improves the longitudinal phase space (upper plot). More specifically, an enhanced laser heater set-point from 0 to 2.5 keV results in an increase of the peak current from about 2 (blue curve, middle plot) to almost 5 kA (yellow curve, middle plot). In this process, one can see that the slice energy spread is also reduced from about 3.5 to 2.5 MeV (bottom plot). As further increasing the induced energy spread by the laser heater, a drop in the peak current and an increase in the slice energy spread can be observed. This is illustrated in Fig.\ \ref{s2e_uB2} in more detail. As shown, the ratio between the peak current (upper plot in Fig.\ \ref{s2e_uB2}) and the central slice energy spread (middle plot in Fig.\ \ref{s2e_uB2}) is scanned over a large range of the laser heater set-point. As a result of the detailed scan, the optimum set-point of the laser heater is 2.5 keV, which corresponds to a minimized ratio (bottom plot in Fig.\ \ref{s2e_uB2}) indicating a maximum achievable peak current and a smallest energy spread at the same time. It can be also noted that further increasing the laser heater set-point from 2.5 keV results in a switch of the working regime from micro-bunching dominated to laser heater dominated. That is, the bunch slice energy spread increases and the peak current drops as increasing the laser heater set-point. The determined optimum ratio between peak current and the associated slice energy spread from the very detailed micro-bunching studies are then considered as the reference to adjusting the set-point of the laser heater model in the following full S2E simulations.
\subsection{Particle Tracking to the SASE Undulator Entrance} \label{tracking} 

Particle tracking simulations are performed in a tracking code OCELOT \cite[]{ocelot}. The collective effects including 3D space charge forces, 3D wake fields and 1D coherent synchrotron radiation are taken into accounts. The power set-point of the laser heater model is adjusted such that an as-close-as-possible ratio between the bunch peak current and the central slice energy spread is obtained after the last bunch compression stage as presented in Section \ref{dE}.

\begin{figure}[!h]
\centering
\includegraphics[width=14cm,height=10cm]{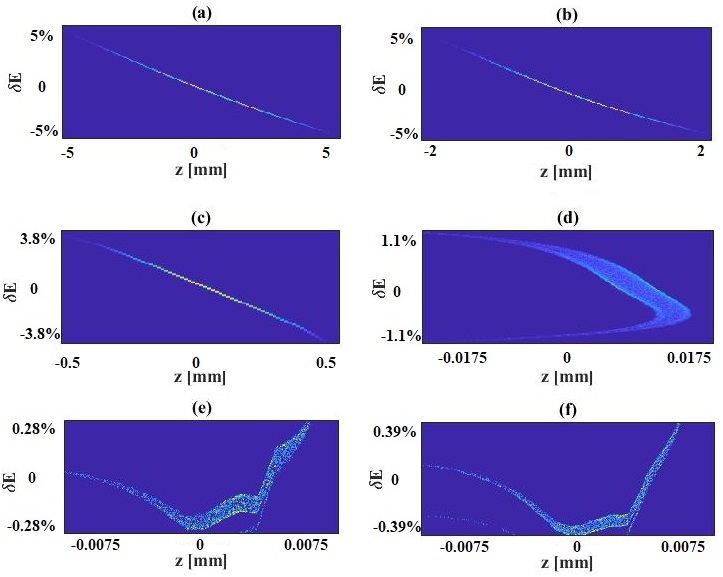}
\caption{Evolution of simulated longitudinal phase spaces along the whole beam line. The vertical axis shows the relative change of the beam energy in percentage while the horizontal axis stands for a relative longitudinal position within the bunch. (a) after injector dogleg; (b) after BC0; (c) after BC1; (d) after BC2; (e) after L3; (f) after CL.}\label{s2e_LPS}
\end{figure}

Simulated longitudinal phase spaces along the beam line are exemplarily shown in Fig.\ \ref{s2e_LPS}. The design beam energies at three bunch compressors are 110, 500 and 2000 MeV, respectively. As seen, a significant energy chirp is applied before the last bunch compressor for necessary compression. This chirp is then partially compensated by the longitudinal Wake fields in the main linac (e). Due to space charge effects an additional energy chirp of alternative sign around the current peak is created in the main linac. In the collimation section, the longitudinal phase space is further impacted by strong wake fields (f). 

\begin{figure}[!h]
\centering
\includegraphics[width=13cm,height=8.5cm]{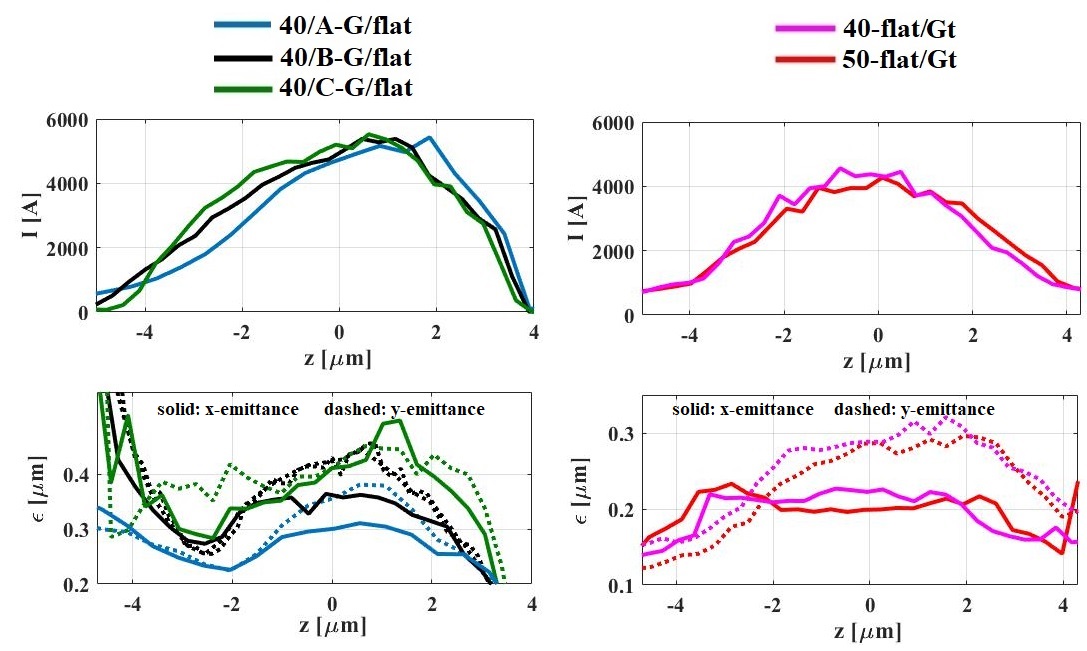}
\caption{Current profile and slice emittance of the simulated particle distributions in front of the SASE undulator beam line using the optimized bunches at the injector. G denotes a Gaussian profile, G$_t$ a truncated Gaussian profile and flat a flattop profile. The numbers in the legend, 40 and 50, denote the gun acceleration gradients.}\label{s2e_caseABCD}
\end{figure}

For case studies, S2E beam dynamics simulations are systematically conducted. The results for the selected bunches from the injector optimization are presented in Fig.\ \ref{s2e_caseABCD}. Note that, for each case the simulated bunch distribution is illustrated at the entrance of the SASE undulator beam line. In Fig.\ \ref{s2e_caseABCD}, the two subplots on the left show the current profiles and slice emittance for the bunches A, B and C at 40 MV/m (blue, black and green curves) as correspondingly listed in Table \ref{Tab:abctab}. The other two subplots on the right show the results for the two bunches with truncated-Gaussian distributions at 40 (pink curve) and 50 MV/m (red curve), respectively.

Table \ref{tab:rf} summarizes the rf parameters found for the S2E studies shown above. Columns 3-6 correspond to the set-points of amplitudes and phases for A1, AH1, L1 and L2, respectively. The L3 set-point is then adjusted for a final beam energy of the design. A summary of beam quality parameters is given in Table \ref{tab:beam} for different bunches used from the injector. The symbols $\epsilon_{xy}$ represent the transverse emittance of the central slice within the bunch. The terms $\textrm I_\textrm{peak}$ and $\delta \textrm E$ stand for the peak current of the bunch and the energy spread of the central slice, respectively. The parameters $\textrm Q_\textrm{core}$ describes charge occupation in percentage within $\pm$ $\sigma_{z}$ around the current peak. As shown in Table \ref{tab:beam}, a peak current of about 5 kA can be achieved for the bunches A, B and C at 40 MV/m, among which the bunch C shows a highest peak current of about 5.5 kA. The lowest central slice emittance is obtained for the bunch at 50 MV/m, which is produced using a longitudinal-flattop and transverse-truncated-Gaussian distribution of the cathode drive laser in the injector. The central slice energy spread varies from 2 to about 2.5 MeV as studied in Sec.\ \ref{dE} in detail. An overall comparison may justify the bunch at a gun acceleration gradient of 50 MV/m as relatively a good choice from the CW injector optimization due to its lowest slice emittance and energy spread. It can be also noted that the central slice emittance is not significantly degraded in comparison to the upstream injector bunches. The emittance growth is mainly due to the compression in the last bunch compressor. This also means, that the choices of the used working points in the S2E simulations of transporting and compressing the bunches are generally feasible.

\begin{table}
	\centering
	\setlength{\tabcolsep}{8pt}
	\caption{Summary of rf parameters used for the different scenarios. G denotes a Gaussian profile, G$_t$ a truncated Gaussian profile and flat a flattop profile.}
	\begin{tabular}{cccccc}
		\hline\hline
		{Gun V$'$/case}& Laser&A1 (A,$\phi$)&AH1 (A,$\phi$)&L1 (A,$\phi$)&L2 (A,$\phi$)\\
		{[MV/m]}&long./trans.&[MV/deg]&[MV/deg]&[MV/deg]&[MV/deg]\\
		\hline
		{40/A}&G/flat	&124.07/3.70   &19.14/162.55     &426.74/23.96   &1515.23/7.93\\
		{40/B} &G/flat	&123.50/0.64   &19.64/155.99  &426.78/23.97  &1516.71/8.27\\
		{40/C} &G/flat   &123.07/-1.65 &19.99/150.88  &426.70/23.94 &1518.15/8.58\\
		\hline
		{40} &flat/G$_t$    &127.29/12.27 &18.85/182.25  &427.19/24.09 &1514.56/7.65\\
		{50} &flat/G$_t$    &126.19/12.07 &18.87/181.59  &427.28/24.12 &1513.75/7.50\\
	\hline\hline
	\end{tabular}\label{tab:rf}
\end{table}

\begin{table}
	\centering
	\setlength{\tabcolsep}{20pt}
	\caption{Summary of key beam quality properties in front of SASE undulators. G denotes a Gaussian profile, G$_t$ a truncated Gaussian profile and flat a flattop profile.}
	\begin{tabular}{lccccc}
		\hline\hline
		{Gun V$'$/case}& Laser&$\mathrm I_\mathrm{peak}$ &$\epsilon_{xy}$ &$\delta \mathrm{E}$&$\mathrm Q_\mathrm{core}$\\
		{[MV/m]}&long./trans.&[kA]&[$\mu m$]&[MeV]&[$\%$]\\
		\hline
		{40/A} 	& G/flat&5.37  &0.32  &2.46 &81.47\\
		{40/B} 	& G/flat&5.17  &0.38  &2.49 &66.36\\
		{40/C}	& G/flat&5.57  &0.43  &2.51 &63.27\\
		\hline
		{40}	&flat/G$_t$&4.56  &0.26  &2.44 &77.17\\
		{50}	&flat/G$_t$&4.10 &0.24  &2.20 &78.04\\
	\hline\hline
	\end{tabular}\label{tab:beam}
\end{table}

Through start-to-end simulations, one can finally justify the advantages of applying advanced laser pulse profiles in improving the beam qualities before the undulators. As shown in Table \ref{tab:beam}, the emittance of the three bunches A, B and C at 40 MV/m are larger than those of the other two bunches at 40 and 50 MV/m, suggesting a combination of the longitudinal-flattop and transverse-truncated-Gaussian profiles is more beneficial compared to the combination of the longitudinal-Gaussian and transverse-uniform distributions. Note that, applying the truncated-Gaussian laser distribution to the simulations of the XFEL injector in pulsed mode results in an improved agreement between measurement and simulation. Based on currently available infrastructure of the laser laboratory at the XFEL, a cathode drive laser system with required shaping possibilities can be further improved and proof-of-principle experiments have been done \cite[]{Test2019}. Taking the advantage of the cathode laser pulse shaping capability, boosting the gun gradient to 50 MV/m can further improve the bunch qualities, i.e. a transverse emittance of about 0.24 $\mu$m is obtained, as shown in Table \ref{tab:beam}. Such a bunch can be compressed to about 4.1 kA with a central slice energy spread of less than 2.5 MeV.

\subsection{Evaluation of Bunch Qualities in FEL Simulations}\label{SASE}
SASE simulations are conducted in Genesis in order to further evaluate the bunch qualities in terms of the lasing performance in hard X-ray regime. The obtained electron bunches, as presented in Table \ref{tab:beam}, are used for studies. The final beam energy is set as about 7.3 GeV according to the cooling capabilities of the current cryogenic system at the European XFEL. In the following, the lasing capability with the optimized bunches is demonstrated in the sub-nanometer and sub-angstrom regimes. The chosen photon beam energies include 9.3 keV (i.e. 0.13 nm), 15 keV (i.e. 0.8 angstrom) and 20 keV (i.e. 0.6 angstrom). Due to technical limits at XFEL, a minimum average $\beta$ function of 15 m in the undulators is set \cite[]{Saldin2004}.

\begin{figure}[!h]
\centering
\includegraphics[width=12cm,height=7.5cm]{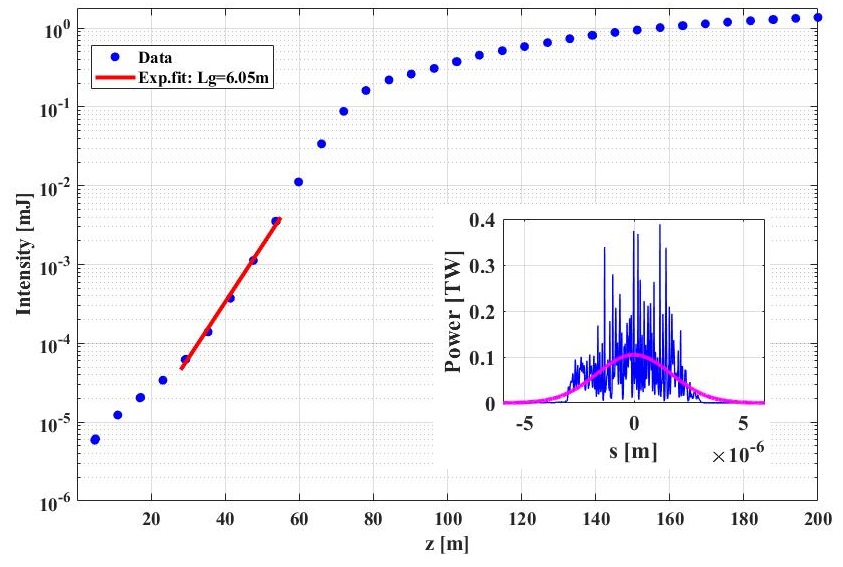}
\caption{SASE pulse energy along the undulators using the electron bunch at 50 MV/m as shown in Table \ref{tab:beam} at a photon energy of 9.3 keV. An undulator period of 4 cm is applied. The pink curve is a Gaussian fit to the power distribution shown in the inset. Note that the gain length is about 4.96 m after excluding the impact of an intersection length of about 1.1 m between neighbouring undulator cells.}\label{lasing9}
\end{figure}

As a first example, the baseline undulators with a period of 4 cm presently in operation at the European XFEL are used for lasing studies. This is for justifying the qualities of the optimized electron bunches taking the advantage that no modification is required to the existing undulators. Figure \ref{lasing9} shows the SASE pulse energy along the baseline undulator beam line for a photon energy of 9.3 keV using the electron bunch distribution at 50 MV/m as shown in Table \ref{tab:beam}. The inset of Fig.\ \ref{lasing9} illustrates the corresponding power distribution by the end of the undulator line. As shown, the optimized electron bunch, characterized by a central slice emittance of about 0.24 $\mu$m, slice energy spread of about 2.2 MeV and a peak current of 4.1 kA, delivers SASE pulses with an intensity above 1 mJ. The red line shows an exponential fit to the gain curve, resulting in a fitted gain length of about 4.96 m. Note that the impact of an intersection length of about 1.1 m between the neighbouring undulator cells has been excluded from the exponentially-fitted gain length as shown in the legend of Fig.\ \ref{lasing9}. 

\begin{figure}[!h]
\centering
\includegraphics[width=12cm,height=7.5cm]{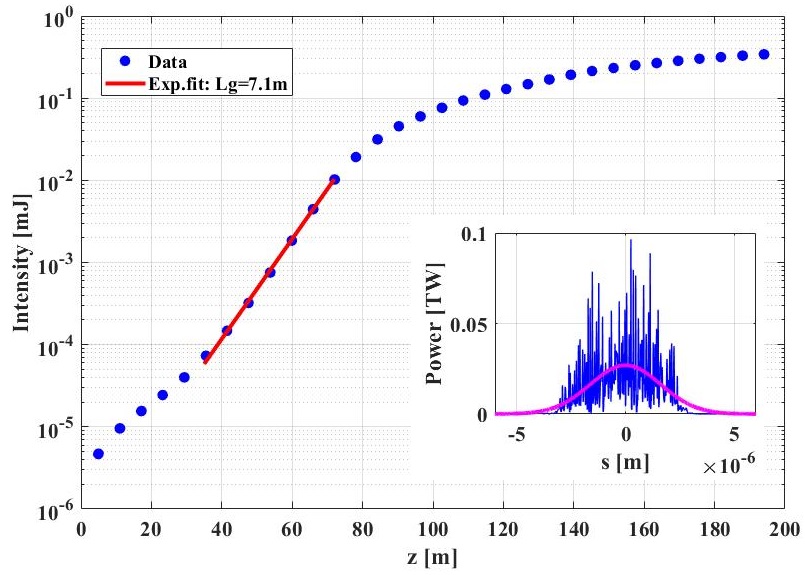}
\caption{SASE signals in sub-angstrom radiation regime (20 keV) using the electron bunch at 50 MV/m as shown in Table \ref{tab:beam}. An undulator period of 2 cm is applied. Note that the gain length is about 5.82 m after excluding the impact of an intersection length of about 1.1 m between neighbouring undulator cells. }\label{lasing10}
\end{figure}

Then, a reduced undulator period of 2 cm is considered as a straightforward approach to access even higher photon energies for compensating a reduced final beam energy for CW operation, as suggested in Eq.\ (\ref{eqlambda}). Figure \ref{lasing10} shows the lasing signal in the sub-angstrom regime at a photon energy of 20 keV using the same electron bunch distribution. As shown, the SASE intensity can reach an order of half millijoule by adjusting the linear and post-saturation taper settings. The resulting gain length in this case is about 5.82 m.

Based on Figs.\ \ref{lasing9} and \ref{lasing10}, it can be noted that the optimized electron bunches from the S2E simulations have shown the capabilities to lase in hard X-ray regime using both baseline undulators and the undulators with a shorter period. The photon energy can reach 20 keV with about half-millijoule SASE intensity. This is consistent with the theoretical prediction given in Fig.\ \ref{undulatorK1} of Sec.\ \ref{intro}, where an undulator period reduction to 2 cm allows lasing at 20 keV and above for an electron beam energy of about 7 GeV at a reasonable gain length below 7 m. 

In addition to Figs. \ref{lasing9} and \ref{lasing10}, more results obtained from systematic lasing studies at the three wavelengths above-mentioned are summarized in Table \ref{tab:fel}. As a reference, an additional comparison of SASE performance at 20 keV between two operation modes (CW and Pulsed) of the design is presented in the bottom block of Table \ref{tab:fel}. As shown, the two bunches with transverse-truncated-Gaussian profiles, as described in Table \ref{tab:beam}, both can lase at 20 keV using a gain length below 7 m. The bunch with a higher gun operating gradient of 50 MV/m delivers a higher photon pulse energy at the sub-angstrom wavelength.

\begin{table}[!h]
	\centering
	\setlength{\tabcolsep}{20pt}
	\caption{Summary of lasing studies at the three wavelengths of about 0.13 nm, 0.08 nm and 0.06 nm for the optimized electron bunches as shown in Table \ref{tab:beam}. The electron beam energy is fixed at 7.3 GeV. An undulator period of 2 cm is considered (except for the case in pulsed mode). A minimum $\beta$ function of 15 m is set. An exemplary comparison of SASE performance between CW mode and Pulsed mode is given at 20 keV for 7.3 GeV and 14 GeV electron bunches, respectively. G denotes a Gaussian profile, G$_t$ a truncated Gaussian profile and flat a flattop profile.}
	\begin{minipage}{\textwidth}
	\centering
	\begin{tabular}{lccccc}
		\hline\hline
        &&$\lambda_\mathrm{p}$ = 0.13 nm (9.3 keV)\\
        \hline
        {Gun V$'$/case}& Laser&$\mathrm{I}$& ${\mathrm P_{m}}$\footnote{peak of a Gaussian fit to the power distribution by the end of the undulator beam line, as illustrated by the insets of Figs.\ \ref{lasing9} and \ref{lasing10}} &$\mathrm L_\mathrm{g}$\footnote{exponential-fitted gain length to gain curves; the impact of an intersection length of 1.1 m between neighbouring undulator cells has been extracted for the numbers shown in this column.}\\
        {[MV/m]}&long./trans.& [mJ]&[W] &[m]\\
        \hline
		{40/A}& G/flat&0.84  &6.2e10  &2.16\\
		{40/B}& G/flat&0.87  &7.1e10  &2.73\\
		{40/C}& G/flat&0.82  &5.9e10  &2.56\\
		{50}&flat/G$_t$&1.09  &7.5e10  &1.95\\
        \hline\hline
        &&$\lambda_\mathrm{p}$ = 0.08 nm (15 keV)&\\
        \hline
        {Gun V$'$/case}& Laser&$\mathrm{I}$& ${\mathrm P_{m}}$&$\mathrm L_\mathrm{g}$\\
        {[MV/m]}&long./trans.& [mJ]&[W] &[m]\\
        \hline
		{40/A}& G/flat&0.38 &3.0e10  &3.73\\
		{40/B}& G/flat&0.37 &3.5e10  &4.83\\
		{40/C}& G/flat&0.35 &3.7e10  &4.82\\
		{50}&flat/G$_t$&0.78 &5.2e10  &3.15\\
		\hline\hline
		&&$\lambda_\mathrm{p}$ = 0.06 nm (20 keV)&\\
		\hline
		{Gun V$'$/case}& Laser&$\mathrm{I}$& ${\mathrm P_{m}}$ &$\mathrm L_\mathrm{g}$\\
	    {[MV/m]}&long./trans.& [mJ]&[W] &[m]\\
        \hline
	    {CW\footnote{CW operation mode using the bunch with a transverse-truncated-Gaussian profile at 40 MV/m in Table \ref{tab:beam} for 100 pC at 7.3 GeV electron beam energy using 2 cm (period) undulators}, 40}&flat/G$_t$&0.3 &2.4e10  &6.83\\
	    
	    {CW\footnote{CW operation mode using the bunch at 50 MV/m in Table \ref{tab:beam} for 100 pC at 7.3 GeV electron beam energy using 2 cm (period) undulators}, 50}&flat/G$_t$&0.5 &2.7e10  &5.82\\
	    
		{Pulsed\footnote{Pulsed operation mode using a bunch with longitudinal-Gaussian and transverse-truncated-Gaussian profiles for 250 pC at 14 GeV electron beam energy using 4 cm (period) undulators.}, 57}&G/G$_t$&2.8  &1.8e11  &5.47\\
	\hline
	\end{tabular}\label{tab:fel}
	\end{minipage}
\end{table}

\subsection{Conclusion and Outlook}\label{Conc}
We present beam parameters and lasing performance for the CW operation of the European XFEL. Our chosen machine lay out aims for photon energies up to about 20 keV with minimum changes to the facility. It requires a cw electron gun, modified rf modules in the injector and the linacs L1 and L2 of the bunch compression system, an additional rf system and new undulator magnets as well as additional cooling capacity. The big linac L3 remains untouched. 

As cw gun, a super-conducting L-band gun is proposed, which could use the present injector lay out, replacing the pulsed gun. Prototypes of such a gun have demonstrated rf field gradients of 40 MV/m with Q$_0\ge10^{10}$ and even 50 MV/m without field emission.
The modified rf modules have been successfully tested at DESY and the new sc undulators with a period length between 15 and 18 mm are planned for an SCU afterburner at EuXFEL.

We have shown that a sc L-band gun driven by a radial uniform laser pulse can be optimized to deliver bunches with a central transverse slice emittance of 0.2 $\mu$rad.
With a truncated Gaussian transverse laser profile, which, by and large, we already apply in the pulsed gun, 0.15 $\mu$rad is possible, and, more demanding, increasing the gradient to 50 MV/m  yields 0.1 $\mu$rad. 

For the transport to the undulator entrance we took great care to treat all collective effects properly. The impedance budget for the wake fields includes all relevant accelerator components, the space charge forces are calculated in full 3D.
The 1D CSR model has been checked with a 2D model to be accurate enough and intra-beam scattering effects are shadowed by the modelled laser heater.

The impact of the micro-bunching instability is calculated by tracking the real number of particles in the bunch, showing the necessity of a laser heater and that slice energy spread significantly below 2.5 MeV cannot be achieved. 

At the entrance of the undulator then, with a peak current of about 5 kA, the central slice emittance for 40 MV/m gun gradient with radial uniform laser pulse is about 0.3 $\mu$rad. For the truncated Gaussian it is about 0.25 $\mu$rad for 40 MV/m gradient and 0.2 $\mu$rad for 50 MV/m.

SASE simulations have then been carried out with the code GENESIS, adjusting linear and quadratic undulator gap tapering to achieve maximum photon pulse energy.
All cases lase in an undulator of 2 cm wave length at 9.3 keV with about 1 mJ photon pulse energy, at 15 keV the 50 MV/m case has twice the power (0.8 mJ compared to 0.4 mJ) and at 20 keV, only the bunches with smaller emittances generated with truncated Gaussian laser profiles at 40 and 50 MV/m  lase,  using gain lengths close to the maximal allowed 7 m to achieve 0.3 and 0.5 mJ pulse energy, respectively.

For comparison: at 14 GeV electron beam energy, lasing in an undulator with 4 cm wave length reaches an energy of 2.8 mJ at 20 keV photon energy with a bunch charge of 0.25 nC. The comparison is fair since the requirements on transverse emittance are easier to meet at higher electron beam energies.  

Lower photon energies between 1 and 10 keV can be reached with an undulator of 4 cm wave length, either with the present XFEL undulators or by applying an sc undulator type under development at the European XFEL, which allows period doubling, for example to switch between 2 and 4 cm period electrically \cite[]{undulator1}. Fig. \ref{lasing9} shows lasing in a 4 cm wave length undulator with an electron beam energy of 7.3 GeV at a photon energy of 9.3 keV with an intensity of 1.5 mJ.

If we stay within the design beam power limit of the pulsed machine of 600 kW, cw operation can provide electron bunches with 0.1 nC charge with a rate of up to 500 kHz, lasing with energies of around 1 mJ at 10 keV and 0.5 mJ at 20 keV.

The integrated photon energy per second at 20 keV of the pulsed machine is about 75 J, cw operation, with the scenario above, reaches about 250 J. This advantage of the cw operation gets more pronounced towards lower photon energies. Integrated photon energy for the cw operation can be further increased by raising the bunch frequency and/or, at lower photon energies, the bunch charge. Both measures require raising the allowed electron beam power level , requiring more beam dumps, and put higher demands on laser power ($\overline{P}\approx$ 5 W at 1 MHz and 0.1 nC at 245 nm on cathode) and thermal load on the cathode.

In total, cw operation delivers lasing with photon energies up to 20 keV with larger photon bunch spacing (its original motivation) and higher integrated photon numbers. The pulsed machine with more than twice the electron beam energy can reach higher photon intensity per pulse and ultra-high photon energies up to about 100 keV.

\subsection{Acknowledgements}\label{Ack}
We like to thank our colleagues R. Brinkmann, S. Casalbuoni, W. Decking, H. Qian, E. Schneidmiller, J. Sekutowicz, S. Tomin and I. Zagorodnov for helpful discussions. Special thanks to E. Gjonaj and S.A.Schmid, TEMF Darmstadt, for the contribution of beam distributions for the Micro-Bunching Instability studies. Special thanks to Ji Qiang, LBNL Berkeley, for the support with IMPACT-Z calculations.

\FloatBarrier

\appendix

\begin{thebibliography}{99}

\bibitem[\protect\citeauthoryear{R. Brinkmann et al.}{2014}]{Brink2014}
R. Brinkmann, E.A. Schneidmiller, J. Sekutowicz, M.V. Yurkov, Nuclear Instruments and Methods in Physics Research Section A: Accelerators, Spectrometers, Detectors and Associated Equipment
Volume 768, 21 December 2014, Pages 20-25
\bibitem[\protect\citeauthoryear{EXFEL TDR}{2006}]{TDR2006}
The European X-Ray Free-Electron Laser Technical design report, DESY 2006-097

\bibitem[\protect\citeauthoryear{Schneidmiller et al.}{2012}]{SYHL2012}
E.A. Schneidmiller, M.V. Yurkov,  Physical Review ST-AB, 15 (2012), p. 080702

\bibitem[\protect\citeauthoryear{Feng et al.}{2013}]{Feng2013}
Guangyao Feng, Igor Zagorodnov, Torsten Limberg, Hyunchang Jin, Yauhen Kot, Martin Dohlus, Winfried Decking, TESLA-FEL 2013-04

\bibitem[\protect\citeauthoryear{Ming-Xie}{1996}]{MX1996}
Ming Xie, 1996, DESIGN OPTIMIZATION FOR AN X-RAY FREE ELECTRON LASER DRIVEN BY SLAC LINAC

\bibitem[\protect\citeauthoryear{J. Sekutowicz et al.}{2005}]{Seku2005} J.K. Sekutowicz, Superconducting RF photoinjectors; an overview ,  \emph{Proc.  The Physics and Applications of High Brightness Electron Beams”},
Workshop, October 9-14, 2005, Erice Italy

\bibitem[\protect\citeauthoryear{J. Sekutowicz et al.}{2007}]{Seku2007}
J.K. Sekutowicz et al. ,  Status of Nb-Pb superconducting RF-gun cavities, 
\emph{Proc. PAC'07}, , Albuquerque, NM, USA, June 25-29, 2007

\bibitem[\protect\citeauthoryear{E. Vogel et al.}{2018}]{Vogel2018} E. Vogel et al. , SRF gun development at DESY,  Proceed.
in \emph{Proc. LINAC'18}, Beijing, China, Sep. 2018, pp. 105--108


\bibitem[\protect\citeauthoryear{E. Vogel et al.}{2019}]{vogel2019} E. Vogel et al. , Status of the All Superconducting Gun Cavity at DESY,  Proceed.
in \emph{Proc. SRF'19}, Dresden, Germany, Jun.-Jul. 2019, pp. 1087--1090


\bibitem[\protect\citeauthoryear{C. F.  Papadopoulos et al.}{2014}]{Popad14}Longitudinal and transverse optimization for a high
repetition rate injector ,  Proceed. in \emph{Proc. FEL'14},  Basel, Switzerland, paper THP057 

\bibitem[\protect\citeauthoryear{URL}{}]{cst}CST MWS, \url{https://www.3ds.com/products-services/simulia/products/cst-studio-suite/}

\bibitem[\protect\citeauthoryear{K.
Floettmann}{}]{astra} K. Floettmann,  A Space Charge Tracking Algorithm, \url{https://www.desy.de/~mpyflo/}

\bibitem[\protect\citeauthoryear{FMM}{}]{FMM} 
    S.A. Schmid, H. De Gersem, E. Gjonaj, TEMF, TU Darmstadt, Darmstadt, Germany, M. Dohlus, DESY, Hamburg, Germany, 
Simulating Space Charge Dominated Beam Dynamics Using FMM, \url{https://doi.org/10.18429/JACoW-NAPAC2019-WEPLE10}

\bibitem[\protect\citeauthoryear{Birte van der Horst, et al.}{2021}]{KEK}
Birte van der Horst, Daniel Klinke, Andrea Muhs, Manuela Schmökel, Jacek
Sekutowicz, Sven Sievers, Nicolai Steinhau-Kühl, Alexey Sulimov, Jan-
Hendrik Thie, Lennart Trelle, Elmar Vogel, "Development and Adjustment of
Tools for Superconducting RF Gun Cavities", Proceedings of 2021
International Conference of RF Superconductivity (SRF'21), 28 June 2021 to 2
July 2021

\bibitem[\protect\citeauthoryear{Kalyanmoy Deb, et al.}{2002}]{nsga2}Kalyanmoy Deb et al.,  A Fast and Elitist Multiobjective Genetic Algorithm: NSGA-II,  \emph{IEEE transactions on evolutionary computation} , vol. 6, no. 2, April 2002

\bibitem[\protect\citeauthoryear{F. Zhou et al.}{2012}]{tgs1}F. Zhou et al., Impact of the spatial laser distribution on photocathode gun operation,  \emph{Phys. Rev. ST Accel. Beams 15, 090701, 2012}

\bibitem[\protect\citeauthoryear{L. Winkelmann}{}]{lutzlaser}
Private communication with Lutz Winkelmann

\bibitem[\protect\citeauthoryear{D. Dowell, J. Schmerge}{2009}]{SLAC}
David H. Dowell and John F. Schmerge, Phys. Rev. Accel. Beams 12, 074201 (2009)
\bibitem[\protect\citeauthoryear{J. Smedley et al.}{2008}]{BNL}
J. Smedley, T. Rao, J. Sekutowicz, Phys. Rev. Accel. Beams 11, 013502 (2008)
\bibitem[\protect\citeauthoryear{R. Barday et al.}{2013}]{HZB} 
R. Barday, A. Burrill, A. Jankowiak et al., Phys. Rev. Accel. Beams 16, 123402 (2013)
\bibitem[\protect\citeauthoryear{E. Saldin et al.}{2004}]{LH1}
E. Saldin, E.Schneidmiller and M.Yurkov, 
Nuclear Instruments and Methods in Physics Research Section A: Volume 528, Issues 1–2, 1 August 2004, Pages 355-359
\bibitem[\protect\citeauthoryear{OCELOT}{}]{ocelot}
OCELOT, https://github.com/ocelot-collab
\bibitem[\protect\citeauthoryear{I. Zagorodnov et al.}{2019}]{Zagorodnov-PRAB2019}
I. Zagorodnov et al., Physical Review Accel. Beams 22 (2019) 024401
\bibitem[\protect\citeauthoryear{W. Decking et al.}{2020}]{XFEL}
W. Decking et al., Nature Photonics 14, 650 (2020)
\bibitem[\protect\citeauthoryear{I. Zagorodnov, M. Dohlus}{2011}]{Compression1}
I. Zagorodnov and M. Dohlus, Physical Review ST Accel. Beams 14, 014403 (2011)
\bibitem[\protect\citeauthoryear{T. Limberg et al.}{2005}]{Compression2}
T. Limberg et al., PAC2005, 1236-1238
\bibitem[\protect\citeauthoryear{M. Dohlus, T. Limberg}{2005}]{Compression3}
 M. Dohlus and T. Limberg, PAC2005, 1015-1017
\bibitem[\protect\citeauthoryear{J. Qiang et al.}{2000}]{impactz1}
J. Qiang, R. Ryne, S. Habib et al., J. Comp. Phys. vol. 163, 434 (2000); IMPACT code available at https://amac.lbl.gov/~jiqiang/IMPACT/
\bibitem[\protect\citeauthoryear{J. Qiang et al.}{2006}]{impactz2}
J. Qiang, S. Lidia, and R. D. Ryne et al., Physical Review Accel. Beams 9, 044204 (2006).
\bibitem[\protect\citeauthoryear{M. Gross et al.}{2019}]{Test2019}
M. Gross et al., in Proc. IPAC’19, Melbourne, Australia, May 2019, pp.1958-1960. doi:10.18429/JACoW-IPAC2019-TUPTS012
\bibitem[\protect\citeauthoryear{S. Casalbuoni et al. 1}{2018}]{Sara}
S. Casalbuoni et al., Synchr. Rad. News, 31:3, 24-28, 2018
\bibitem[\protect\citeauthoryear{S. Casalbuoni et al. 2}{2019}]{undulator1}
S. Casalbuoni et al., 2019 J. Phys.: Conf. Ser. 1350 012024
\bibitem[\protect\citeauthoryear{S. Casalbuoni et al.}{2021}]{Sara1}
S. Casalbuoni et al., WEPAB132 IPAC2021

\bibitem[\protect\citeauthoryear{S. Abeghyan et al.}{2019}]{p_undulator1}
S. Abeghyan et al., J. Synchrotron Rad. 26, pp. 302-310, 2019


\bibitem[\protect\citeauthoryear{E.L.Saldin et al.}{2004}]{Saldin2004}
E.L.Saldin, E.A.Schneidmiller, M.V.Yurkov, Nuclear Instruments and Methods in Physics Research Section A: Accelerators, Spectrometers, Detectors and Associated Equipment, Volume 235, Issues 4–6, 15 May 2004, Pages 415-420

\end{thebibliography}
\end{document}